\documentclass[fleqn,usenatbib]{mnras}
\usepackage[dvipdfmx]{graphicx}
\usepackage{pdfpages}

\usepackage{amsmath}	
\usepackage{amssymb}	

\usepackage{bm}
\usepackage{multirow}

\title[Dust coagulation]{Dust coagulation and fragmentation in a collapsing cloud core and their influence on non-ideal magnetohydrodynamic effects}
\author[Y.  Kawasaki et al.]{
Yoshihiro Kawasaki,$^{1}$\thanks{E-mail: kawasaki.yoshihiro.592@s.kyushu-u.ac.jp (YK)} 
Shunta Koga,$^{1}$ and 
Masahiro N. Machida$^{1}$
\\
$^{1}$Department of Earth and Planetary Sciences, Faculty of Sciences, Kyushu University, Fukuoka 819-0395, Japan
}

\begin{document}
\label{firstpage}
\pagerange{\pageref{firstpage}--\pageref{lastpage}}
\maketitle

\begin{abstract}
We determine the time evolution of the dust particle size distribution during the collapse of a cloud core, 
accounting for both dust coagulation and dust fragmentation, to investigate the influence of dust growth on non-ideal magnetohydrodynamic effects.
The density evolution of the collapsing core is given by a one-zone model. 
We assume two types of dust model: dust composed only of silicate (silicate dust) and dust with a surface covered by $\mathrm{H_{2}O}$ ice ($\mathrm{H_{2}O}$ ice dust). 
When only considering collisional coagulation, the non-ideal magnetohydrodynamic effects are not effective in the high-density region for both the silicate 
and $\mathrm{H_{2}O}$ ice dust cases. 
This is because dust coagulation reduces the abundance of small dust particles, resulting in less efficient adsorption of charged particles on the dust surface.
For the silicate dust case, when collisional fragmentation is included, the non-ideal magnetohydrodynamic effects do apply at a high density 
of  $n_{\mathrm{H}}>10^{12} \ \mathrm{cm^{-3}}$ because of the abundant production of small dust particles.  
On the other hand, for  the $\mathrm{H_{2}O}$ ice dust case, the production of small dust particles due to fragmentation is not efficient.
Therefore, for  the $\mathrm{H_{2}O}$ ice dust case, non-ideal magnetohydrodynamic effects apply only in the range $n_{\mathrm{H}}\gtrsim 10^{14} \ \mathrm{cm^{-3}}$, 
even when collisional fragmentation is considered. 
Our results suggest that it is necessary to consider both dust collisional coagulation and fragmentation to activate non-ideal magnetohydrodynamic 
effects, which should play a significant role in the star and disk formation processes.
  
\end{abstract}



\section{Introduction}
Magnetic fields play an important role in star and disk formation processes \citep{2012PTEP.2012aA307I,2018FrASS...5...39W,2020SSRv..216...43Z}. 
During star formation in a collapsing cloud core, 
angular momentum should be removed from the collapsing core by  magnetic effects  such as magnetic braking 
\citep{1994ApJ...432..720B,1995ApJ...452..386B,1995ApJ...453..271B,2011MNRAS.413.2767M,2011PASJ...63..555M,2000ApJ...528L..41T,2002ApJ...575..306T} 
and magnetically driven winds \citep{1982MNRAS.199..883B,1985PASJ...37..515U,2000prpl.conf..759K,2000ApJ...528L..41T}.
In the star formation process, non-ideal magnetohydrodynamic (MHD) effects (Ohmic dissipation, ambipolar diffusion, and the Hall effect) arising from the weakly ionized plasma of the collapsing cloud core \citep{1999MNRAS.303..239W,2002ApJ...573..199N} play an important role in the  evolution of the magnetic field.
Both Ohmic dissipation and ambipolar diffusion allow the first core to evolve into a circumstellar disk \citep{2011MNRAS.413.2767M}.
The dissipation of the magnetic field suppresses  excessive angular momentum transport by magnetic braking and promotes the formation of a rotationally supported (or circumstellar) disk 
\citep{2013ApJ...763....6T,2015ApJ...801..117T,2010A&A...521L..56D,2012A&A...541A..35D,2010ApJ...724.1006M,2015MNRAS.452..278T,2021MNRAS.502.4911X,2021MNRAS.508.2142X}.
The Hall effect determines the size of the rotationally supported disk formed in the collapsing cloud core depending on whether the angular momentum vector of the initial core is aligned with the direction of  the magnetic field.  
When the angular momentum vector is parallel to the direction of the magnetic field,  a small disk forms, 
whereas a large disk appears when  the angular momentum vector is anti-parallel to the direction of the magnetic field  \citep{2015ApJ...810L..26T,2021MNRAS.507.2354W}.

Dust has a strong effect on the magnetic diffusion coefficients, which determine the strength of  the non-ideal MHD effects.
The magnetic diffusion coefficients are determined by the quantity of charged particles. 
Dust greatly reduces the abundance of charged particles, as the charged particles are adsorbed on the dust surface.
 In many studies, dust-related reactions have been  incorporated into chemical reaction networks \citep{1990MNRAS.243..103U,2002ApJ...573..199N,2016A&A...592A..18M,2016MNRAS.460.2050Z,2018MNRAS.478.2723Z,2019MNRAS.484.2119K}.
These studies considered different dust particle size distributions that can affect the number of charged particles and the magnetic diffusion coefficients. 
In three-dimensional non-ideal MHD calculations, different dust particle size distribution models produce different outcomes in the star and disk formation processes
\citep{2018MNRAS.473.4868Z,2021MNRAS.505.5142Z,2020ApJ...896..158T}.
Therefore, the dust particle size distribution is an important factor in star and disk formation processes.

Many studies assume that the dust particle size does not evolve when calculating the magnetic diffusion coefficients and non-ideal MHD effects. 
In other words, the dust particle size (distribution) was fixed in past studies \citep{2009ApJ...693.1895K, 2016MNRAS.460.2050Z,2018MNRAS.478.2723Z,2019MNRAS.484.2119K,2021MNRAS.501.5873W}. 
However, when dust particles collide with each other, they coagulate or fragment. 
Thus, the dust particle size distribution should evolve with time
and it is important to understand the evolution to determine the effect on the magnetic diffusion coefficients.
In almost all three-dimensional simulations, the set of magnetic diffusion coefficients is prepared in advance by performing chemical reaction calculations assuming a certain dust distribution.
Then, during the simulation,  the magnetic diffusion coefficients at each point (or particle) are extrapolated using the physical properties at each point (or particle), such as density, temperature, and magnetic field strength
\citep[e.g.,][]{2018MNRAS.473.3080M}.

Recently, two-fluid (dust and gas) calculations have been performed \citep{2019A&A...626A..96L,2020A&A...641A.112L,2021ApJ...913..148T}, though the computational cost of this approach is very high. 
In addition, it is difficult to incorporate the evolution of the dust particle size distribution in such studies. 
Evolution of the dust particle size distribution in collapsing cloud cores has typically been evaluated using one-zone or one-dimensional calculations instead of three-dimensional simulations \citep{2009MNRAS.399.1795H,2020A&A...643A..17G,2021A&A...649A..50M}. 
\citet{2020A&A...643A..17G} calculated the evolution of the dust particle size distribution in a contracting molecular cloud core 
and determined the evolution of the magnetic diffusion coefficients.
In their study, an MRN distribution \citep{1977ApJ...217..425M} was adopted as the initial size distribution, in which small dust particles were removed by coagulation, 
resulting in a significant change in the magnetic diffusion coefficients.
However, they only considered coagulation growth of dust particles and did not consider fragmentation. 
Fragmentation due to collisions between dust particles produce dust particles smaller than the colliding dust particles, 
which can significantly affect the dust particle size distribution evolution.

In this study, we calculate the evolution of the dust particle size distribution in a collapsing cloud core, 
taking into account both dust coagulation and dust fragmentation.
The density evolution is given by a one-zone model.
From the size distribution evolution, we estimate the ionization degree and magnetic diffusion coefficients to  evaluate the non-ideal MHD effects. 

This paper is structured as follows.
We describe the basic equations, coagulation and fragmentation models, and calculation method of the diffusion coefficients in Section~\ref{sec:method}.
The results are presented in Section~\ref{sec:results}.
The implications for star and disk formation and caveats are discussed in Section~\ref{sec:discussion}. 
A summary is presented  in Section~\ref{sec:summay}.

\section{Methods}
\label{sec:method}
\subsection{Basic equation}
In a collapsing cloud core, the dust particle size distribution changes due to coagulation and fragmentation caused by the collision of dust particles.
The time variation of the dust mass density $\rho \left(m, t\right)$ for a dust mass $m$ at time $t$ is expressed by the following coagulation-fragmentation equation
\citep{1916ZPhy...17..557S,2018MNRAS.475..167B}.
\begin{align}
  \frac{\mathrm{d}\rho\left(m,t\right)}{\mathrm{d}t}
  & = \frac{1}{2}\int_{0}^{m} m K\left(m-m_{1}, m_{1}\right)n\left(m-m_{1}, t\right) \notag \\ 
  & \ \ \ \ \times n\left(m_{1}, t\right) \mathrm{d}m_{1}\notag \\
  & - \int_{0}^{\infty}mK\left(m,m_{1}\right)n\left(m, t\right)n\left(m_{1}, t\right) \mathrm{d}m_{1} \notag \\
  & + \frac{1}{2}\int \int_{0}^{\infty} m F\left(m_{1}, m_{2}\right) n\left(m_{1}, t\right) n\left(m_{1}, t\right) \notag \\ 
  & \ \ \ \ \times \varphi_{\rm f}\left(m ; m_{1}, m_{2}\right) 
  \mathrm{d}m_{1}\mathrm{d}m_{2} \notag \\ 
  & - \int_{0}^{\infty} m F\left(m,m_{1}\right)n\left(m, t\right)n\left(m_{1}, t\right) \mathrm{d}m_{1} \notag \\
  & + \frac{\rho\left(m, t\right)}{\rho_{g}}\frac{\mathrm{d}\rho_{g}}{\mathrm{d} t},
  \label{eq:coag-frag-equation}
\end{align}
where $n\left(m, t\right)$ is the number density of dust mass $m$, $K\left(m, m_{1}\right)$ and $F\left(m, m_{1}\right)$ are the collisional and fragmentation kernels,
$\varphi_{f} \left(m; m_{1}, m_{2}\right)$ is the distribution function for fragments after a collision between $m_{1}$ and $m_{2}$ dust particles, and $\rho_{g}$ is the gas mass density.

The first two terms on the right-hand side of Eq.~(\ref{eq:coag-frag-equation}) correspond to an increase or decrease in dust density due to coagulation.
The next two terms represent an increase or decrease in dust density due to fragmentation.
The last term represents the change in the dust distribution due to the change in ambient gas density \citep{2019MNRAS.482.2555H}.
In this study, the last term corresponds to the density change of the collapsing cloud core. 

It should be noted that, in addition to the effects of coagulation and fragmentation, the dust particle size can increase as the ambient gas accretes onto the dust surface.
Although this study does not take into account the increase in dust particle size due to this effect,  its impact is addressed in Section~\ref{sec:discussion}.
Moreover, each dust particle is assumed to be compact and spherical in this study, while actual dust particles may be distorted or have porosity.
We will focus on these cases in our subsequent studies.

\subsection{Relative velocity}
\label{sec:relavel}

Dust motion is an important factor in determining the collision rate with respect to dust growth.
As the origin of the relative velocity between dust particles, we consider both turbulence and thermal motion  (or Brownian motion).
In this subsection, after we simply explain our turbulence model, we describe thermal motion. 

Gas turbulence is a key factor in determining dust motion because dust interacts with gas in the collapsing cloud core.
In this study, we adopt the model presented by \citet{2007A&A...466..413O} for the relative velocity between dust particles induced by turbulence.
We assume the turbulence at any scale larger than the viscous scale of turbulence and consider it within the Jeans scale at a given cloud density.
Although we do not need to strictly identify the origin of the turbulence, we assume that the turbulence at any scale arises from multi-scale cascade of bulk-flow fluctuations in a molecular cloud core. 
In addition, we also assume that the turbulence obeys  the  Kolmogorov law and the amplitude (or  fluctuating velocity) at the Jeans scale corresponds to the sound speed for convenience. 
It should be noted that there exists non-negligible turbulence motion in star-forming cores \citep[e.g.][]{1981MNRAS.194..809L}, while  the turbulence velocity  comparable to the sound speed, which is adopted in this and previous studies,  may be a bit large (see also \S\ref{sec:turbulence}).
We will focus on the effect of turbulence intensity on dust growth in our forthcoming paper.
In the following, we describe the prescription to relate the turbulence to the relative velocity between dust particles. 

We define the turnover time of the largest turbulence eddy with the Jeans length as the sound crossing time,
\begin{equation}
  \tau_{\mathrm{L}} = \frac{L_{\mathrm{J}}}{c_{s}} = \frac{1}{2}\sqrt{\frac{\pi}{G\rho_{g}}}.
\end{equation}
As described above, the fluctuating velocity is given by the sound speed $v_{\mathrm{t}} = c_{s}$,  where $c_{s} = \sqrt{k_{B}T/m_{\mu}}$, $k_{B}$ is Boltzmann constant, $T$ is the gas temperature, and $m_{\mu}$ is the mean molecular mass of the gas.
With a Kolmogorov turbulent cascade, the eddy-turnover time of the turbulent motions at the viscous scale can be describe as
\begin{equation}
  \tau_{\eta} = \frac{\tau_{\mathrm{L}}}{\sqrt{\mathrm{Re}}},
\end{equation}
where $\mathrm{Re} = \nu_{t}/\nu_m$ is the the Reynolds number.
The turbulent viscosity $\nu_{t}$  is described as $\nu_{t}=L v_{\mathrm{t}} = L_{J} c_{s}$. 
 $\nu_{m}$ is the molecular viscosity.
Thus,  the Reynolds number is describe as \citep{2009A&A...502..845O}
\begin{equation}
  \mathrm{Re} = \frac{\nu_{t}}{\nu_{m}} = 6.2\times 10^{7} \left(\frac{\rho_{g}/m_{\mu}}{10^{5} \ \mathrm{cm^{-3}}}\right)^{1/2} \left(\frac{T}{10 \ \mathrm{K}}\right)^{1/2},
\end{equation}

Dust dynamics in turbulent flow is controlled by the stopping time. The stopping time for dust based on the Epstein law is described as
\begin{equation}
  \label{eq:epstein_stopping_time}
  \tau = \frac{\rho_{s} a}{\rho_{g} v_{\mathrm{th}}},
\end{equation}
where $\rho_{s}$, $a$, and $v_{\rm th}$ are the dust internal density, dust particle radius, and thermal velocity of the molecular gas, 
defined as $v_{\mathrm{th}} = \sqrt{8/\pi}c_{s}$, respectively.
The Epstein law holds when the relationship $a < (9/4)l_{g}$ is realized between the dust radius $a$ and the mean free path $l_{g}$ for the gas.
If $a > (9/4)l_{g}$, the Epstein law is not applicable and the Stokes law can be applied. 
The stopping time for the Stokes law is described as
\begin{equation}
  \tau = \frac{4 a^{2}}{9 l_{g} v_{\mathrm{th}} \rho_{g}}.
\end{equation}
All dust particle sizes in this study are within  the range of the Epstein law.

When considering the relative velocity due to turbulence of two dust particles, we use $\tau_{i}$ for the particle with the larger stopping time and $\tau_{j}$ for the smaller stopping time.
The relative velocity due to Kolmogorov turbulence is given in three regions as follows \citep{2007A&A...466..413O},
\begin{equation}
  \Delta V^{\mathrm{T}}_{ij} = 
  \begin{cases}
    \sqrt{\frac{3}{2}} c_{s} \sqrt{\frac{\tau_{i} - \tau_{j}}{\tau_{i}+ \tau_{j}}}
      \left[ \frac{ \left( \tau_{i}/\tau_{\mathrm{L}} \right)^{2} }{\tau_{i}/\tau_{\mathrm{L}}+ \mathrm{Re}^{-1/2}}
        - \frac{ \left( \tau_{j}/\tau_{\mathrm{L}} \right)^{2} }{\tau_{j}/\tau_{\mathrm{L}}+ \mathrm{Re}^{-1/2}} \right]^{1/2} \\ 
      \ \ \ \ \ \ \ \ \ \ \ \ \ \ \ \ \ \ \ \ \ \ \ \ \ \ \ \ \ \ \ \ \ \ \ \ \ \ \ \ \  (\tau_{i} < \tau_{\eta}), \\ 
      \sqrt{\frac{3}{2}} c_{s} \sqrt{ f\left(\frac{\tau_{\mathrm{j}}}{\tau_{\mathrm{i}}}\right) \frac{\tau_{\mathrm{i}}}{\tau_{\mathrm{L}}} }
      \ \ \ \ \ \ \ \ \ \ \ \ \ \ \ \ \ \ (\tau_{\eta} \le \tau_{i} < \tau_{\mathrm{L}}), \\ 
      \sqrt{\frac{3}{2}} c_{s} \sqrt{ \left( \frac{1}{1 + \tau_{\mathrm{i}/\tau_{\mathrm{L}}}} + \frac{1}{1 + \tau_{\mathrm{j}/\tau_{\mathrm{L}}} } \right) }
      \ \ \ \  (\tau_{\mathrm{L}} < \tau_{i}),
  \end{cases}
\end{equation}
where the function $f$ can be described as 
\begin{equation}
  f\left(x\right) = 3.2 - \left(1 + x\right) + \frac{2}{1 + x} \left( \frac{1}{2.6} + \frac{x^{3}}{1.6 + x} \right).
\end{equation}

In addition to the relative velocity due to turbulence, the contribution from thermal motion of the dust  (Brownian motion) is also considered:
\begin{equation}
  \Delta V^{\mathrm{B}}_{ij} = \sqrt{\frac{8 k_{B} T (m_{i} + m_{j}) }{\pi m_{i} m_{j}}}.
\end{equation}
Thermal motion contributes mainly to the motion of dust particles with sizes smaller than $0.1 \ \mathrm{\mu m}$.
Thus, the total relative velocity can be expressed as
\begin{equation}
  \label{eq:rel_vel_total}
  \Delta V_{ij} = \sqrt{ \left(\Delta V^{\mathrm{T}}_{ij} \right)^{2} + \left( \Delta V^{\mathrm{B}}_{ij} \right)^{2} }.
\end{equation}

\subsection{Coagulation and fragmentation kernel}
The outcome of dust collision depends on various factors such as the relative velocity between dust particles, 
the collision cross section, and the internal properties of the dust.
Although they can all be included in the kernel, it is difficult to deal with all of them, and many studies have simplified the problem.
In this study, we use a dust coagulation and fragmentation kernel that takes into account the probabilistic distribution of dust relative velocities
\citep{2013ApJ...764..146G,2018MNRAS.475..167B}.
The coagulation $K_{ij}$ and fragmentation $F_{ij}$ kernels of dust particles $i$ and $j$ are described as
\begin{align}
  & K_{ij} = s_{ij} \int_{0}^{\infty} \Delta v P_{ij}(\Delta v) \epsilon^{\mathrm{c}}_{ij} \left(\Delta v \right) \mathrm{d} \Delta v, \label{eq:coag_kernel_1} \\ 
  & F_{ij} = s_{ij} \int_{0}^{\infty} \Delta v P_{ij}(\Delta v) \epsilon^{\mathrm{f}}_{ij} \left(\Delta v \right) \mathrm{d} \Delta v, \label{eq:frag_kernel_1} 
\end{align}
where $s_{ij}$ is the collision cross section and is estimated as $s_{ij} = \pi \left( a_{i} + a_{j} \right)^{2}$ assuming each dust particle is a sphere with radii $a_{i}$ and $a_{j}$.

$P_{ij}\left(\Delta v\right)$ is the probability distribution function for the relative velocity of two dust particles.
The collision velocity between dust particles of arbitrary size is assumed to be given by a Gaussian distribution with random motions induced by 
both Brownian motion and turbulence as the variance \citep{2013ApJ...764..146G}:
\begin{equation}
  P_{ij} \left(\Delta v\right) = \sqrt{\frac{2}{\pi}} \frac{\Delta v^{2}}{\sigma^{3}_{ij}}\exp\left(-\frac{\Delta v^{2}}{2\sigma^{2}_{ij}}\right),
\end{equation}
where $\Delta v$ is the relative velocity of the dust and $\sigma_{ij}$ is the variance,
\begin{equation}
  \sigma^{2}_{ij} = \frac{\pi}{8} \left[ \left(\Delta V^{\mathrm{T}}_{ij} \right)^{2} + \left( \Delta V^{\mathrm{B}}_{ij} \right)^{2} \right]
         = \frac{\pi}{8} \Delta V_{ij}^{2}.
\end{equation}

In equations (\ref{eq:coag_kernel_1}) and (\ref{eq:frag_kernel_1}), $\epsilon^{c}_{ij}\left(\Delta v\right)$ and $\epsilon^{f}_{ij}\left(\Delta v\right)$ 
represent the probability of coagulation or fragmentation of dust particles after a collision, respectively.
These probabilities depend on the collision velocity and the properties of the dust.
However, in this study, we simply assume that the dust particles coagulate when the relative velocity between dust particles is $v<v_{c}$ and fragments when $v>v_{f}$
\citep{2012A&A...544L..16W}.
In other words, $\epsilon^{c}_{ij}\left(\Delta v\right)$ and $\epsilon^{f}_{ij}\left(\Delta v\right)$ are given by a Heaviside step function as follows,
\begin{align}
  & \epsilon^{c} \left(\Delta v\right) = H \left(v_{c} - \Delta v\right), \\ 
  & \epsilon^{f} \left(\Delta v\right) = H \left(\Delta v - v_{f} \right),
\end{align}
where  $v_{c}$ and $v_{f}$ are introduced in Section~\ref{sec:fragmentation_model}.

The coagulation and fragmentation kernels can then be expressed as follows,
\begin{align}
  \label{eq:coagulation_kernel}
  K_{ij} &= s_{ij} \int_{0}^{\infty} \Delta v P_{ij}(\Delta v) \epsilon^{\mathrm{c}} \left(\Delta v \right) \mathrm{d} \Delta v \notag \\
         &= s_{ij} \int_{0}^{v_{c}} \Delta v P_{ij}(\Delta v) \mathrm{d} \Delta v \notag  \\ 
         &= s_{ij} \Delta V \left[ 1 - \left( 1 + \frac{v_{c}^{2}}{2\sigma^{2}_{ij}} \right) \exp \left(-\frac{v_{c}^{2}}{2\sigma^{2}_{ij}}\right) \right],
\end{align}

\begin{align}
  \label{eq:fragmentation_kernel}
  F_{ij} &= s_{ij} \int_{0}^{\infty} \Delta v P_{ij}(\Delta v) \epsilon^{\mathrm{f}} \left(\Delta v \right) \mathrm{d} \Delta v \notag \\
         &= s_{ij} \int_{v_{f}}^{\infty} \Delta v P_{ij}(\Delta v) \mathrm{d} \Delta v \notag  \\ 
         &= s_{ij} \Delta V \left( 1 + \frac{v_{f}^{2}}{2\sigma^{2}_{ij}} \right) \exp \left(-\frac{v_{f}^{2}}{2\sigma^{2}_{ij}}\right).
\end{align}

In this study, we calculate the charge state of  dust particles to estimate the magnetic diffusion coefficients of the non-ideal MHD effects (see \S~\ref{sec:calc_ionization} and \S\ref{appendix:MRN_no_size_evolution}). 
However, we do not consider the charge of dust particles  in the coagulation and fragmentation kernel (eqs.~(\ref{eq:coagulation_kernel}) and (\ref{eq:fragmentation_kernel})). 
Dust charging influences  the collision cross section \citep{1993A&A...280..617O}.
For example, dust charging can act as a repulsive force for dust particles charged with the same sign when  the collisional energy of the two dust particles is less than their Coulomb energy \citep{2009ApJ...698.1122O}.
Dust charging also influences  the relative velocity between dust particles because the Lorentz force acts on the charged dusts \citep{2020A&A...643A..17G}.
We will investigate the effects of charged dust particles on the evolution of the dust size distribution in a future study.

\subsection{Fragmentation model}
\label{sec:fragmentation_model}
When dust fragmentation occurs as a result of the collision of two dust particles $\left( m_{1}>m_{2} \right)$, 
the fragments obey the following mass distribution function $\varphi_{\rm f}$,
\begin{equation}
  m\varphi_{\rm f}(m; m_{1}, m_{2}) = m_{\rm rm}\delta\left(m - m_{\rm m}\right)
    + m g_{\rm f}\left(m; m_{1}, m_{2}\right) .
\end{equation}
The first term on the right-hand side represents the dominant mass remaining in the fragments, and the second term represents the continuous distribution of the other masses.
The continuous mass distribution function is represented by the power distribution $g_{\rm f} \propto m^{-\xi}$ in the range $(m_{\rm f, min} \le m \le m_{\rm f, max})$.
The conservation of mass before and after the collision can be described as  
\begin{equation}
  \label{eq:mass_conservation}
  m_{1} + m_{2} = \int_{0}^{\infty} m \varphi_{\rm f}(m; m_{1}, m_{2}) \ \mathrm{d}m. 
\end{equation}
The normalization constant of $g_{\mathrm{f}}$ can be obtained from equation (\ref{eq:mass_conservation}), and $g_{f}$ is
\begin{equation}
  g_{\rm f}\left(m; m_{1}, m_{2}\right) = \frac{\left(m_{1}+m_{2}-m_{\rm rm}\right) \left(2-\xi\right)}
    {m_{\rm f, max}^{2-\xi}-m_{\rm f, min}^{2-\xi}} m^{-\xi}.
  \label{eq:frag_mass_cont_dis}
\end{equation}
In this study, we use $\xi = 11/6 = 1.83$ \citep{2008A&A...480..859B,2010Icar..206..735K}.

We need $m_{\mathrm{rm}}$ to determine the mass distribution function.
We use the relational equation obtained from numerical dust collision experiments to determine the mass $m_{\mathrm{ej}}$ ejected after the two dust particles collide
\citep{2013A&A...559A..62W},
\begin{equation}
  m_{\rm ej} = \frac{v}{v_{\rm col, crit}} m_{2},
  \label{eq:eject_mass}
\end{equation}
where $v$ is the collision velocity for dust particles and $v_{\mathrm{col,crit}}$ is the velocity required for most of the dust particles to fragment after the collision.
For $v_{\mathrm{col,crit}}$, \citet{2013A&A...559A..62W} obtained the following scaling relation:
\begin{equation}
  v_{\mathrm{col, crit}} \simeq 20 \sqrt{\frac{E_{\mathrm{break}}}{m}},
  \label{eq:u_col_crit}
\end{equation}
where $E_{\mathrm{break}}\simeq 23 \left[ \gamma^{5}a^{4} \left( 1- \nu^{2}\right)^{2} / \mathcal{E}^{2}  \right]^{1/3}$ is the energy required to break contact between two particles of radius $a$, 
$\gamma$ is the surface energy of the particles, $\mathcal{E}$ is Young's modulus, and $\nu$ is Poisson's ratio.
Thus, $m_{\mathrm{rm}}$ is described as
\begin{equation}
  m_{\rm rm} = m_{1} + m_{2} - m_{\rm ej}.
\end{equation}
When the collision velocity $v$ is less than $v_{\mathrm{col,crit}}$, $m_{ej}$ is less than $m_{2}$. In this case, $m_{\mathrm{rm}}$ is greater than $m_{1}$. 
In other words, the fragmentation model used in this study also includes mass transport between the two colliding dust particles.

The equations obtained in \citet{2013A&A...559A..62W} consider the case of dust composed of a single monomer.
However, there is no single-size monomer in interstellar dust, and there is instead a size distribution \citep{1977ApJ...217..425M}.
In this study, we incorporate the effect of the size distribution in $u_{\mathrm{col, crit}}$ in the following simple way based on \citet{2009A&A...502..845O}.
$E_{\mathrm{break}}/m$ represents the total binding energy per unit mass, or strength of the substance.
If the dust is composed of monomers with a single size, then $E_{\mathrm{break}}/m \propto a^{-5/3}$, 
which indicates that the smaller the monomer size the greater the dust strength.
When considering the size distribution of monomers, we assume that the contact between monomers always contains a small monomer,
and $E_{\mathrm{break}}$ is evaluated with the minimum monomer size $a_{\mathrm{min}}$.
Furthermore, assuming that the number of contacts is of the same order of magnitude as the number of monomers comprising the dust, 
the average strength is given by
\begin{align}
  \frac{ \langle E_{\mathrm{break}}\rangle }{\langle m \rangle}
  & = \frac{ \int_{a_{\mathrm{min}}}^{a_{\mathrm{max}}} E_{\mathrm{break}} \left(a\right) \frac{\mathrm{d}n}{\mathrm{d}a} \mathrm{d}a  }
    { \int_{a_{\mathrm{min}}}^{a_{\mathrm{max}}} m\left(a\right) \frac{\mathrm{d}n}{\mathrm{d}a} \mathrm{d}a } \notag \\ 
  & \simeq \frac{ E_{\mathrm{break}} \left(a_{\mathrm{min}}\right) \int_{a_{\mathrm{min}}}^{a_{\mathrm{max}}} a^{-q} \mathrm{d}a  }
  { \frac{4}{3}\pi\rho_{s} \int_{a_{\mathrm{min}}}^{a_{\mathrm{max}}} a^{3-q} \mathrm{d}a } \notag \\
  & = \frac{ E_{\mathrm{break}} \left(a_{\mathrm{amin}}\right)  }{ m_{\mathrm{min}} }
    \frac{ \left(4-q\right) \left(\eta^{1-q} - 1\right) }{\left(1-q\right) \left(\eta^{4-q} - 1\right)},
  \label{eq:average_dust_strength}
\end{align}
where $m_{\mathrm{min}}=(4/3)\pi\rho_{s}a_{\mathrm{min}}^{3}$ and $\eta = a_{\mathrm{max}}/a_{\mathrm{min}}$.
Using this average strength, $v_{\mathrm{col, crit}}$ can be described as
\begin{align}
  v_{\mathrm{col, crit}} &= 20 \sqrt{ \frac{\langle E_{\mathrm{break}}\rangle}{\langle m \rangle}} \notag \\ 
                         &\simeq  20 \sqrt{ \frac{\langle E_{\mathrm{break} }\left(a_\mathrm{min}\right)\rangle}{\langle m_{\mathrm{min}} \rangle}} 
                            \sqrt{ \frac{ \left(4-q\right) \left(\eta^{1-q} - 1\right) }{\left(1-q\right) \left(\eta^{4-q} - 1\right)} } \notag \\ 
                        &\simeq v_{0} \left( \frac{a_{\mathrm{min}}}{0.1 \ \mathrm{\mu m}} \right)^{-5/6}
                          \sqrt{ \frac{ \left(4-q\right) \left(\eta^{1-q} - 1\right) }{\left(1-q\right) \left(\eta^{4-q} - 1\right)} }.
\end{align}
The constant value of $v_{0}$ is determined by the physical properties of the dust,
and is estimated as $v_{0} \simeq 80\, \mathrm{ms^{-1}}$ for $\mathrm{H_{2}O}$ ice and $v_{0} \simeq 8\, \mathrm{m s^{-1}}$ for silicate.
In this study, we use $v_{\mathrm{col, crit}}$ to estimate the ejected mass $m_{\mathrm{ej}}$ in equation~(\ref{eq:eject_mass}).
As the collision velocity decreases relative to $v_{\mathrm{col, crit}}$, the ejected mass becomes smaller and deviates from the relation given
in equation~($\ref{eq:eject_mass}$) \citep{2013A&A...559A..62W}. 
Thus, we set the upper velocity limit for coagulation $v_{c}$ and the lower velocity limit for fragmentation $v_{f}$ as $v_{c} = v_{f} = 0.2\, v_{\mathrm{col, crit}}$.
For the MRN size distribution \citep{1977ApJ...217..425M}, which has the parameters $q=3.5$, $a_{\mathrm{min}}=0.005\, \mathrm{\mu m}$, and $a_{\mathrm{max}}=0.25\, \mathrm{\mu m}$, 
the critical collision velocity $v_{\mathrm{col, crit}}$ and the upper velocity limit for coagulation $v_{c}$ are 
$v_{\mathrm{col, crit}}\simeq 180\,  \mathrm{ms^{-1}}$ and  $v_{c} \simeq 35\, \mathrm{ms^{-1}}$ for $\mathrm{H_{2}O}$ ice
and $v_{\mathrm{col, crit}}\simeq 18 \ \mathrm{ms^{-1}}$, $v_{c} \simeq 3.5\, \mathrm{ ms^{-1}}$ for silicate. 

The minimum mass $m_{\rm f, min}$ of the mass distribution function $g_{f}\left(m; m_{1}, m_{2}\right)$ is set to agree with 
the minimum mass $m_{\rm min}$ of the initial size distribution.
The maximum mass $m_{\rm f, max}$ is set to be $m_{\rm f, max} = 0.1\, m_{\rm ej}$.
If the collision velocity $u$ is greater than $v_{\rm col, crit}$,  most of the dust particle is destroyed and $m_{\rm f, max} > m_{\rm rm}$ is realized. 
In this case, we do not consider $m_{\rm rm}$ $(=0)$, and assume that the mass of $m_{\rm ej} = m_1+m_2$ is distributed 
according to the distribution function $g_{f}\left(m; m_{1}, m_{2}\right)$ (eq.~\ref{eq:frag_mass_cont_dis}).

\begin{figure}
  \includegraphics[width=\linewidth]{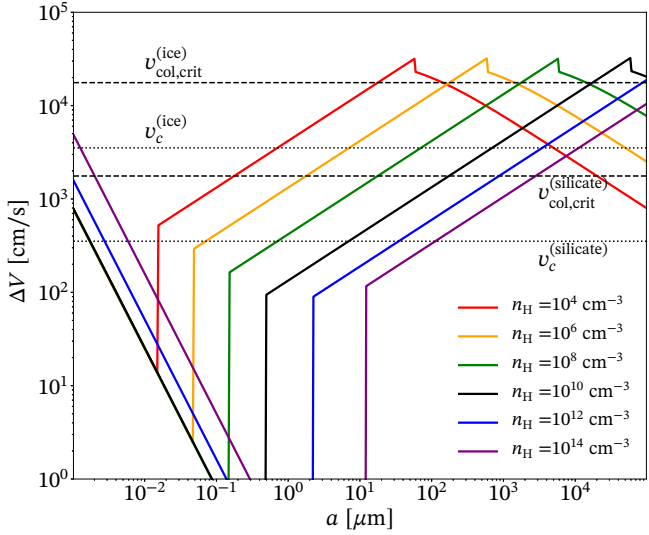}
  \caption{Relative velocity between two same sized dust particles as function of dust particle size (eq.~[\ref{eq:rel_vel_total}]).
  The critical collision velocity $v_{\mathrm{col, crit}}$ and the upper velocity limit for coagulation $v_{c}$ are also plotted for both the silicate and $\mathrm{H_{2}O}$ ice cases.}
  \label{fig:rel_vel}
\end{figure}

Figure~\ref{fig:rel_vel} plots the relative velocity $\Delta V$  between two dust particles against the dust particle size for each gas density (eq.~[\ref{eq:rel_vel_total}]).
The relative velocity is calculated using the same sized dust particles.
The critical collision velocity $v_{\mathrm{col, crit}}$ and the upper velocity limit for coagulation $v_{c}$ are also plotted for both the silicate and $\mathrm{H_{2}O}$ ice cases.
The relative velocity decreases as  the gas density increases. 
Fragmentation of the dust particles begins to occur for collisions where the relative velocity exceeds $v_{c}$.
If the relative velocity is larger than $v_{\mathrm{col, crit}}$, most of the dust particles fragment.
At relatively low densities, fragmentation occurs for smaller dust particle sizes in silicate dust than in $\mathrm{H_{2}O}$ ice dust, as seen in Figure~\ref{fig:rel_vel}.

\subsection{Gas collapse model}
\label{sec:gascollapse}
The time evolution of the gas density due to gravitational collapse of a molecular cloud core is calculated from the one-zone model used in \citet{2005ApJ...626..627O}.
The model assumes a runaway contraction of a molecular cloud core in a self-similar manner.
The gas density evolution is described as
\begin{equation}
  \frac{\mathrm{d}\rho_{g}}{\mathrm{d}t} = \frac{\rho_{g}}{t_{\rm ff}} \sqrt{1 - f},
\end{equation}
where $t_{\rm ff}$ is the free-fall time,
\begin{equation}
  t_{\rm ff} = \sqrt{\frac{3\pi}{32 G\rho_{g}}}.
\end{equation}
$f$ represents the ratio of the magnitude of the pressure gradient force and gravity, 
and is given as a function of the ratio of the specific heat $\gamma = \mathrm{d} \ln P_{g} / \mathrm{d} \ln \rho_{g}$ ($P_{g}$ is gas pressure) as follows:
\begin{equation}
  f = 
  \begin{cases}
    0 \quad \qquad \qquad \qquad  \qquad \qquad \qquad \quad \ \ \ \ \   \left(\gamma < 0.83 \right), \\ 
    0.6 + 2.5 \left( \gamma - 1 \right) - 6.0 \left( \gamma -1 \right)^{2} \quad \quad \quad \left( 0.83 < \gamma < 1\right), \\ 
    1.0 + 0.2 \left(\gamma - 4/3 \right) - 2.9 \left( \gamma - 4/3 \right)^{2} \quad \left( \gamma \ge 1 \right).
  \end{cases}
\end{equation}
The factor $(1-f)^{1/2}$  delays the cloud contraction due to gas pressure.
The factor $f$ becomes  unity when $\gamma = 4/3$, indicating the formation of a (first) hydrostatic core. 
In reality, the first hydrostatic core slowly contracts as envelope gas accretes on to it \citep{2000ApJ...531..350M}.
To model the  slow contraction,  we set $f = 0.95$ with $\gamma \ge 4/3$ according to \citet{2005ApJ...626..627O}.

When considring dust growth, we implicitly introduce turbulence as the origin of the relative velocity between dust particles, as described in \S\ref{sec:relavel}. 
Thus,  the turbulent pressure affects the cloud contraction.
In addition, both magnetic field and rotation can slow the cloud contraction \citep[e.g.][]{2005MNRAS.362..369M}.
These effects should  be included in $f$.
However, we ignore them in this study, because it is very difficult to implement these effects in our one-zone model.

The gas pressure is obtained from the ideal gas equation of state:
\begin{equation}
  P_{g} = \frac{\rho_{g}k_{B}T}{m_{\mu}}.
\end{equation}
In this study, the gas temperature is derived using a barotropic equation of state, used in \citet{2018MNRAS.478.2723Z}:
\begin{equation}
  T = 
  \begin{cases}
    T_{0} + 1.5 \frac{\rho_{g}}{10^{-13}} \ \ \ \ \ \ \ \ \ \ \ \ \ \ \ \  \left( \rho_{g} < 10^{-12} \right), \\ 
    \left( T_{0} + 15 \right) \left(\frac{\rho_{g}}{10^{-12}}\right)^{0.6} \ \ \ \ \ \ \ \  \left( 10^{-12} \le \rho_{g} < 10^{-11} \right), \\ 
    10^{0.6} \left(T_{0} + 15\right) \left(\frac{\rho_{g}}{10^{-11}}\right)^{0.44} \left( 10^{-11} \le \rho_{g} \right),
  \end{cases}
\end{equation}
where $T_{0} = 10 \ \mathrm{K}$ is adopted.

\subsection{Calculation of ionization degree and non-ideal magnetic diffusion coefficients}
\label{sec:calc_ionization}

Dust adsorbs charged particles on its surface, thereby reducing the ionization degree and significantly affecting the magnetic diffusion coefficients \citep{2016A&A...592A..18M,2018MNRAS.478.2723Z,2019MNRAS.484.2119K}. 
As the dust particle size increases, the average dust charge increases \citep{1987ApJ...320..803D} 
and it is difficult to solve a chemical reaction network that includes a large quantity of charged dust particles.
Thus, in this study, we calculate the abundances of electrons $n_{e}$ and ions $n_{i}$ and the average charge number $\langle Z \rangle_{k}$ for each dust particle size using the method proposed 
by \citet{2021A&A...649A..50M}, and estimate the magnetic diffusion coefficients.
With this method, the ionization equilibrium state can be obtained using the analytical solution for the charge distribution of charged dust particles.
The calculation considers the generation of electrons and ions by ionization reactions caused by cosmic rays and other factors, 
and the decrease of electrons and ions by recombination and adsorption on dust.

The physical quantities required for the calculation are as follows:
average ion mass $m_{\rm i} = 25\, m_{\rm P}$, recombination rate of electrons and ions $\langle \sigma v \rangle_{\rm ie} = 2\times 10^{-7} \left( T/300\right)^{-1/2}$, 
and probability coefficients for adsorption of electrons $s_{\rm e} = 0.6$ and ions $s_{\rm i} = 1.0$ on dust. 
The ionization rate $\zeta = \zeta_{\rm CR} + \zeta_{\rm RA}$ accounts for the ionization rate due to cosmic rays $\zeta_{\rm CR}$ 
and the ionization due to radionuclides $\zeta_{\rm RA}$.
Charged particles are mainly produced by the ionization of $\rm H_{2}$ and $\rm He$.
The ionization rate of $\rm He$ is related to that of $\rm H_{2}$ by $\zeta^{\left(\rm He\right)} = 0.84 \zeta^{\left(\rm{H_{2}}\right)}$.
The total ionization rate is given by $\zeta = \zeta^{\left(\rm H_{2}\right)} x_{\rm H_{2}} + \zeta^{\left(\rm He\right)} x_{\rm He}$,
where $x_{\rm H_{2}} = n_{\rm H_{2}}/n_{\rm H}$ and $x_{\rm He} = n_{\rm He}/n_{\rm H}$, in which  $x_{\rm H_{2}}=0.5$ and  $x_{\rm He} = 9.75\times 10^{-2}$ are adopted  \citep{2000ApJ...543..486S}.
Taking attenuation into account, the ionization rate due to cosmic rays is given as follows \citep{1981PASJ...33..617U},
\begin{equation}
  \zeta^{\rm H_{2}} = \zeta_{\rm CR}^{0} \exp\left(-\frac{\Sigma}{\Sigma_{\rm CR}}\right),
\end{equation}
where $\zeta_{\rm CR}^{0} = 1.0\times 10^{-17} \ \mathrm{s^{-1}}$, $\Sigma_{\rm CR} = 96 \ \rm{g cm^{-2}}$ is the attenuation length 
and the average column density $\Sigma$ is  described as \citep{2002ApJ...573..199N}
\begin{equation}
  \Sigma = \sqrt{\frac{k_{B}T\rho_{g}}{\pi G m_{\mu}}}.
\end{equation}
The ionization rate due to radionuclides is adopted as $\zeta_{\rm RA} = 7.3\times 10^{-19} \ \rm s^{-1}$ \citep{2009ApJ...690...69U}.

The magnetic diffusion coefficients for Ohmic $\eta_{O}$, Hall $\eta_{H}$, and ambipolar diffusion $\eta_{A}$ are given by \citep{2007Ap&SS.311...35W,2021MNRAS.504.5588K}
\begin{align}
  \eta_O &= \frac{c^2}{4\pi\sigma_O}, \label{eq:eta_ohm} \\ 
  \eta_H &= \frac{c^2\sigma_H}{4\pi\sigma^2_\perp}, \\
  \eta_A &= \frac{c^2\sigma_P}{4\pi\sigma^2_\perp} - \eta_O,
\end{align}
where $\sigma_{\perp} = \sqrt{\sigma_{H}^{2} + \sigma_{P}^{2}}$, and $\sigma_{O}, \sigma_{H}$, and $\sigma_{P}$
represent the Ohmic, Hall, and Pedersen conductivity, respectively, and are given by
\begin{align}
  \sigma_O &= \sum_{j}\sigma_{O, j} = \frac{c}{B}\sum_{j}Q_j n_j \beta_j, \label{eq:sigma_ohmic} \\
  \sigma_H &= \sum_{j}\sigma_{H, j} =  -\frac{c}{B}\sum_{j}\frac{Q_{j}n_{j}\beta_{j}^{2} }{1+\beta^2_j},  \label{eq:sigma_hall} \\
  \sigma_P &= \sum_{j}\sigma_{P, j} = \frac{c}{B}\sum_{j}\frac{n_{j}Q_{j}\beta_j}{1+\beta^2_j}. \label{eq:sigma_pedersen}
\end{align}
$n_{j}$ represents the number density of charged particles and $Q_{j}$ represents the charge of the charged particles.
$\beta_{j}$ is called the Hall Parameter and represents the relative strength of the Lorentz force acting on charged particles 
and the drag force due to collisions with neutral particles, and is defined as follows,
\begin{equation}
  \beta_{j} = \frac{Q_{j} |\bm{B}|}{m_{j}c} \frac{m_{n} + m_{j}}{\rho_{n} \langle \sigma v \rangle_{j}},
\end{equation}
where $\langle \sigma v \rangle_{j}$ represents the momentum transport efficiency coefficient between neutral and charged particles, and its value is given in \citet{2008A&A...484...17P}.
According to \citet{2002ApJ...573..199N} and \citet{2011ApJ...738..180L}, we adopt the magnetic field as a function of density as  
\begin{equation}
  B = 1.43\times 10^{-7} \sqrt{n_{\rm H}}.
\label{eq:bevo}
\end{equation}
Equation~(\ref{eq:bevo}) can be applicable for $n_{\mathrm{H}} \lesssim 10^{11} \ \mathrm{cm^{-3}}$, while it may not be applicable  for $n_{\mathrm{H}} \gtrsim  10^{11} \ \mathrm{cm^{-3}}$.
In a high-density gas region ($n_{\mathrm{H}} \gtrsim  10^{11} \ \mathrm{cm^{-3}}$), it becomes effective that magnetic field is decoupled from the gas by ambipolar diffusion and Ohmic dissipation \citep{2010MNRAS.408..322K} 
and the amplification of the magnetic field cannot be simply described as a function of density. 
The coefficient of Ohmic dissipation ($\eta_O$) is independent of magnetic field strength, while the Hall ($\eta_{H}$) and ambipolar diffusion ($\eta_{A}$) coefficients depend on  magnetic field strength. 
Since  we adopt  equation~(\ref{eq:bevo})  for all density ranges for simplicity,  we may overestimate the magnetic field strength and the diffusion coefficients of $\eta_{H}$ and $\eta_{A}$ in the range  $n_{\mathrm{H}} \gtrsim 10^{11} \ \mathrm{cm^{-3}}$.

\subsection{Calculation parameters and initial conditions}

The internal density of the dust is $ \rho_{s} = 2.65 \ \rm{g\, cm^{-3}}$ \citep{2009A&A...502..845O}. 
In this study, we consider silicate dust and dust whose surface is covered with $\mathrm{H_{2}O}$ ice.
$\mathrm{H_{2}O}$ ice evaporates when the temperature rises above about 150$\,$K, which is ignored in this study.

Although equation~(\ref{eq:coag-frag-equation}) is written in a continuous form, 
we solve it by discretizing the size distribution.
The computational domain of the size distribution is in the range $a=10^{-3}$--$10^{6}\, \mathrm{\mu m}$, 
and this range is divided into $N_{\mathrm{bin}}=405$ bins. The detailed calculation method is described in Appendix~\ref{appendix:numerical}.

The initial dust particle size distribution is assumed to follow the MRN distribution 
$\mathrm{d}n/\mathrm{d}a \propto a^{-3.5} \left( a_{\mathrm{min, ini}} < a < a_{\mathrm{max, ini}} \right)$.
The minimum size is set to $a_{\mathrm{max,ini}}=5.0\times 10^{-3} \ \mathrm{\mu m}$ and the maximum size is $a_{\mathrm{max}}=2.5\times 10^{-1} \mathrm{\mu m}$.
A mass ratio of gas to dust of $f_{\mathrm{dg}} = 0.01$ is adopted and the dust and gas are assumed to be perfectly coupled. 
In this study, we calculate the dust particle size distribution  during the gravitationally collapsing core phase, and  compare the results with and without fragmentation. 

Table~\ref{tab:model_para} summarizes the parameters for our calculations.
In model sil-coag, silicate dust is considered without fragmentation.
Silicate dust with fragmentation is considered in model sil-frag.
Model ice-coag and ice-frag are the models with and without fragmentation for $\mathrm{H_{2}O}$ dust, respectively.
Note that coagulation of dust particles is considered in all the models listed in Table~\ref{tab:model_para}.
The calculation starts at a number density of $n_{\mathrm{H}}=10^{4} \mathrm{cm^{-3}}$ and ends at $n_{\mathrm{H}}=10^{14} \mathrm{cm^{-3}}$.

\begin{table}
  \caption{
 Model name, dust composition, and whether fragmentation is considered.
  }       
  \begin{tabular}{cccc}
  \hline
  Model            & material   & fragmentation \\ \hline

  sil-coag         & silicate   & No  \\
  sil-frag         & silicate   & Yes \\ 
  ice-coag         & ice        & No \\ 
  ice-frag         & ice        & Yes \\

  \hline
  \end{tabular}
  \label{tab:model_para}
\end{table}

\section{results}
\label{sec:results}

\subsection{Silicate dust case}

\begin{figure*}
  \includegraphics[width=\linewidth]{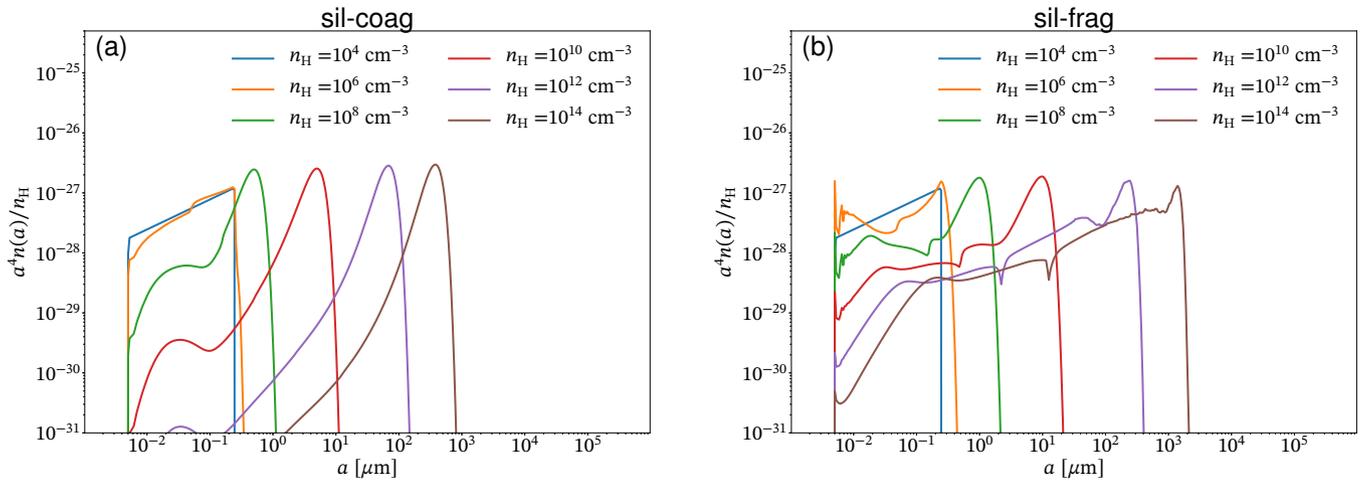}
  \caption{Size distribution evolution for silicate dust. The left panel represents the coagulation only model (sil-coag) and the right panel represents
  the model including fragmentation (sil-frag).}
  \label{fig:sil_size}
\end{figure*}

\begin{figure*}
  \includegraphics[width=140mm]{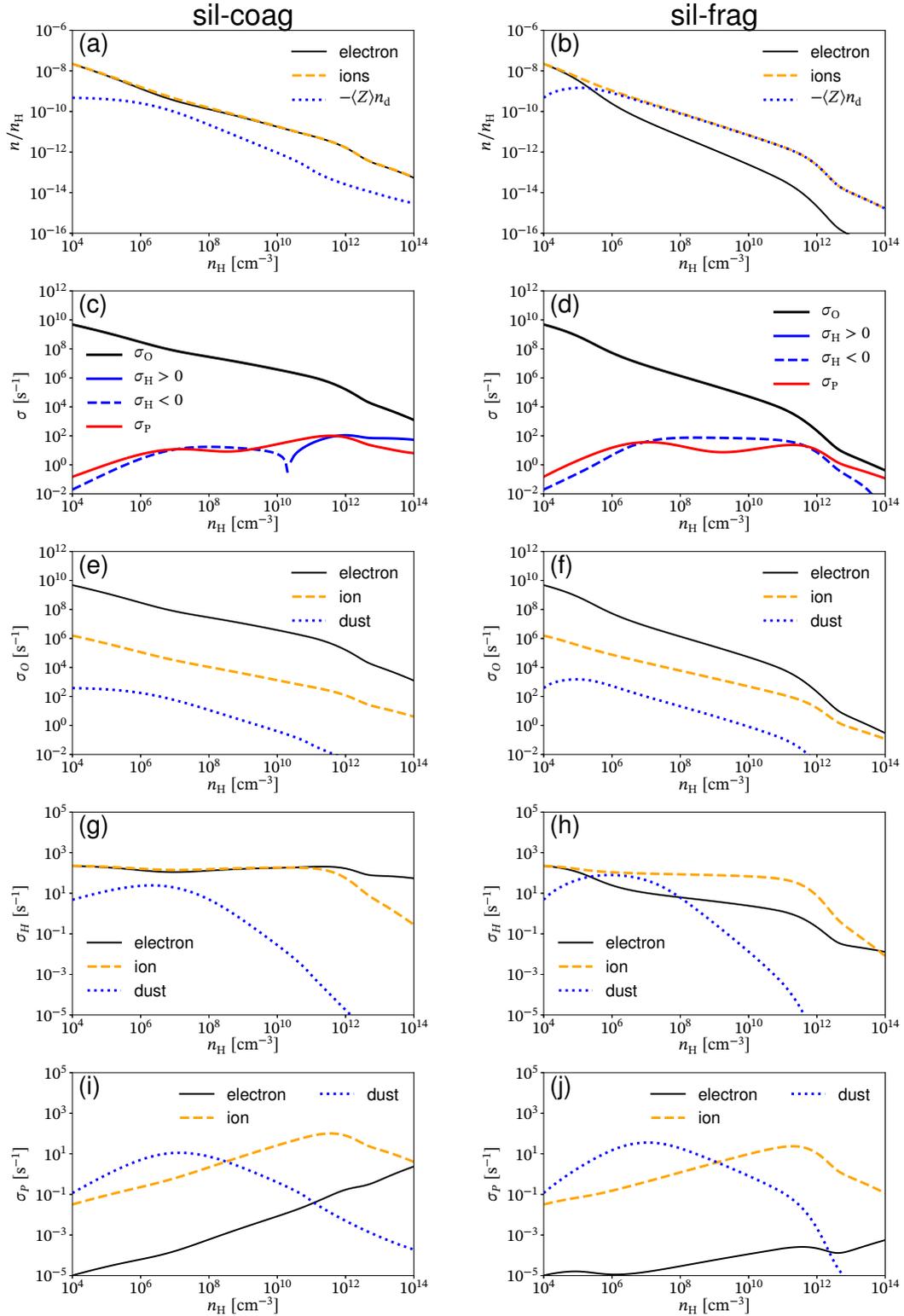}
  \caption{Abundance of charged particles ({\it a}, {\it b}), conductivities ({\it c}, {\it d}),  and decomposition of contributions of electrons, ions, and dust to 
  Ohmic ({\it e}, {\it f}), Hall ({\it g}, {\it h}) and Pedersen ({\it i}, {\it j}) conductivity against number density for models sil-coag (left panels) and sil-frag (right panels). 
  In the top panels, the net dust charge density $\langle Z \rangle n_{\mathrm{d}}$ with a negative sign is also plotted.}
  \label{fig:sil_abund_cond}
\end{figure*}

\begin{figure*}
  \includegraphics[width=140mm]{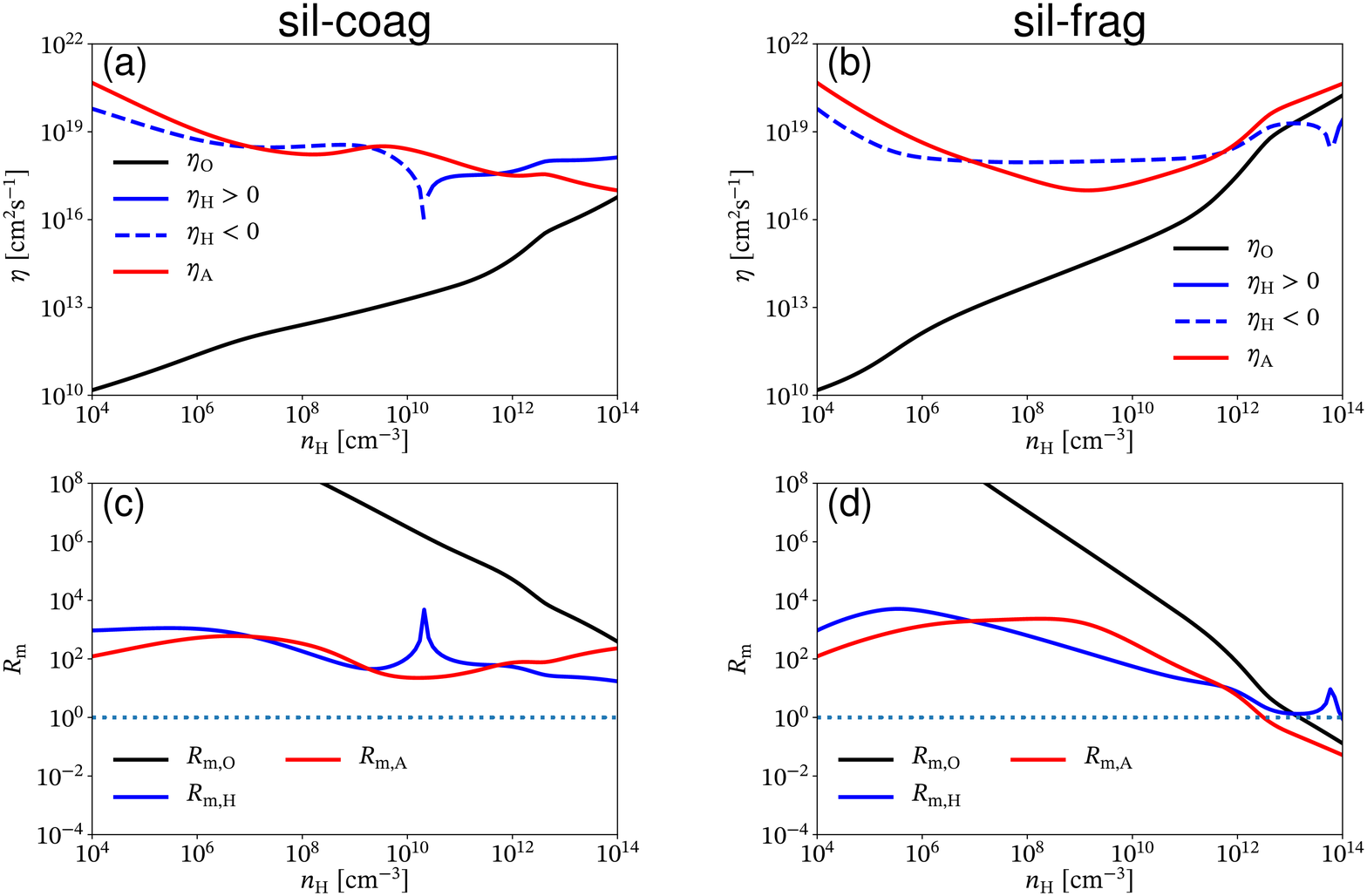}
  \caption{Magnetic diffusion coefficients, $\eta_{O}, \eta_{H}$, and $\eta_{A}$ (top panels) and magnetic Reynolds number $R_{\mathrm{m}}$ (bottom panels) against 
  number density for models sil-coag (left panels) and sil-frag (right panels). 
  The subscripts ``O", ``H", and ``A" indicate Ohmic dissipation, Hall effect, and ambipolar diffusion, respectively.
  The dotted line in the bottom panels corresponds to $R_{\mathrm{m}}=1$.}
  \label{fig:sil_res}
\end{figure*}

Figure~\ref{fig:sil_size} shows the evolution of the dust particle size distribution.
The left panel shows the evolution of the dust particle size distribution for the silicate dust coagulation only model (sil-coag).
At low densities in the early stages of contraction, i.e.\ $n_{\mathrm{H}} < 10^{6} \ \mathrm{cm^{-3}}$, 
the small dust particles coagulate under collision and the number of dust particles with $a \simeq 0.1 \mathrm{\mu m}$ slightly increase.
However, there is little change in the upper limits of the size distribution ($a \simeq 0.25 \mathrm{\mu m}$).
This is because the dust particles with  $a \gtrsim 0.1 \mathrm{\mu m}$ have a relative velocity exceeding the upper velocity limit for coagulation $v_{c}$ (Fig.~\ref{fig:rel_vel}).
For $n_{\mathrm{H}}>10^{8} \mathrm{cm^{-3}}$, the relative velocity between dust particles due to turbulence becomes small, 
and the dust particles with  $a \gtrsim 0.1  \mathrm{\mu m}$ can coagulate and grow to large sizes.
As the density increases, the size distribution becomes concentrated around one size, 
growing to about $a \sim 400 \ \mathrm{\mu m}$ at $n_{\mathrm{H}}=10^{14} \ \mathrm{cm^{-3}}$.

The right panel of Figure~\ref{fig:sil_size} shows the evolution of the dust particle size distribution for the model taking into account fragmentation of silicate dust (sil-frag).
Even at low density, we can confirm that collisional fragmentation increases the number of small sized dust particles compared to model sil-coag. 
The small dust particles with  $a \lesssim 0.1  \mathrm{\mu m}$ are abundant at relatively low densities of $n_{\mathrm{H}} < 10^{8} \mathrm{cm^{-3}}$,  because the increase in small dust particles due to fragmentation dominates  the decrease due to collisional coagulation growth.
The dust particles with $ a \lesssim  0.1  \mathrm{\mu m}$ decrease at higher densities  because the relative velocity becomes small  and thus coagulation growth becomes dominant with less fragmentation occurring.
On the other hand, dust particles with a size of $a > 10  \mathrm{\mu m}$ tend to fragment after a collision, supplying smaller dust particles.
As a result, small dust particles are more abundant in model sil-frag than in model sil-coag.
As noted in Section~\ref{sec:fragmentation_model}, the fragmentation model used in this study also includes mass transport
and thus the maximum size is larger in model sil-frag than in model sil-coag. 
The maximum size of dust particles reaches $a \sim 1000\, \mu m $ at $n_{\mathrm{H}} = 10^{14}\, \mathrm{cm^{-3}}$.

The first  row of Figure~\ref{fig:sil_abund_cond} shows the abundance of ions and electrons and the net dust charge density $\langle Z \rangle n_{\mathrm{d}}$,
where $n_{\mathrm{d}} \equiv \sum_{i} n_{i}$ is the total dust number density, and the average dust change number $\langle Z \rangle$ is defined as
\begin{equation}
  \langle Z \rangle \equiv \frac{\sum_{i} Z_{i} n_{i}}{\sum_{i} n_{i}} = \frac{\sum Z_{i} n_{i}}{ n_{\mathrm{d}}}.
\end{equation}
Dust is net negatively charged because electrons have a greater thermal velocity and collide more frequently with dust than ions.
Note that the figures show the net dust charge density with a negative sign.
For model sil-coag, the abundance of electrons and ions are almost the same at all densities (Fig.~\ref{fig:sil_abund_cond}{\it a}).
Without considering  the coagulation growth of dust, the abundance of electrons in the gas phase is significantly reduced,
because electron adsorption on the dust surface is more efficient than the generation of free electrons by ionization
\citep{1990MNRAS.243..103U,2018MNRAS.478.2723Z,2019MNRAS.484.2119K}.
The abundance of charged particles, conductivities, and magnetic diffusion coefficients without the  coagulation growth of dust are shown  in 
Appendix~\ref{appendix:MRN_no_size_evolution}.
Considering  the coagulation growth of dust, the abundance of electrons increases because the small dust particles decrease 
in abundance due to coagulation growth (Fig.~\ref{fig:sil_size}{\it a}), 
making the adsorption of charged particles (especially electrons) on the dust surface less efficient.
The slight drop in the abundance of charged particles at $n_{\mathrm{H}} = 10^{12} \ \mathrm{cm^{-3}}$ is due to the high density; 
above this value, the attenuation of cosmic rays becomes significant.

When both coagulation and fragmentation are included (model sil-frag), many small dust particles are produced (Fig.~\ref{fig:sil_size}{\it b}),  allowing for efficient capture of electrons. Thus, the abundance of electrons is reduced for $n_{\mathrm{H}}\gtrsim 10^{6} \ \mathrm{cm^{-3}}$ (Fig.~\ref{fig:sil_abund_cond}{\it b}).
As a result, instead of electrons, dust becomes the main carrier of negative charge.
The abundance of ions in model sil-frag is also smaller than that in model sil-coag, especially for $n_{\mathrm{H}} > 10^{12} \ \mathrm{cm^{-3}}$.

The second row of Figure~\ref{fig:sil_abund_cond} shows the conductivities, $\sigma_{O}, \sigma_{H}$ and $\sigma_{P}$.
The third and subsequent rows of Figure~\ref{fig:sil_abund_cond} show decompositions of the contributions of electrons, ions, and dust to 
Ohmic (Fig.~\ref{fig:sil_abund_cond}{\it e} and {\it f}), Hall (Fig.~\ref{fig:sil_abund_cond}{\it g} and {\it h}) 
and Pedersen (Fig.~\ref{fig:sil_abund_cond}{\it i} and {\it j}) conductivity.
For model sil-coag, the Ohmic conductivity is always larger than the other two (Fig.~\ref{fig:sil_abund_cond}{\it c}),
and its value is determined primarily by the electron contribution (Fig.~\ref{fig:sil_abund_cond}{\it e}).
The Hall conductivity is determined by the contribution of electrons and ions (Fig.~\ref{fig:sil_abund_cond}{\it g}).
Note that the Hall conductivity for electrons $\sigma_{H,e}$ is positive by definition (eq.~\ref{eq:sigma_hall}). 
For $n_{\mathrm{H}} < 10^{11} \ \mathrm{cm^{-3}}$, the contribution of ions to the Hall conductivity is slightly larger than that of electrons,
while for $n_{\mathrm{H}} > 10^{11} \ \mathrm{cm^{-3}}$, the contribution of electrons becomes larger (Fig.~\ref{fig:sil_abund_cond}{\it g}).
This corresponds to the change in sign of $\sigma_{H}$  from negative to positive at $n_{\mathrm{H}} \simeq 10^{10} \ \mathrm{cm^{-3}}$ (Fig.~\ref{fig:sil_abund_cond}{\it c}).
Although the abundance of electrons and ions are almost the same for $n_{\mathrm{H}} > 10^{12} \mathrm{cm^{-3}}$ (Fig.~\ref{fig:sil_abund_cond}{\it a}), 
the contribution of ions to the Hall conductivity is smaller than that of electrons (Fig.~\ref{fig:sil_abund_cond}{\it g}). 
This is because the Lorentz force acting on ions is weaker  than the drag force due to collisions with neutral particles.
For Pedersen conductivity, the contribution from dust is largest for $n_{\mathrm{H}} \lesssim 10^{8} \ \mathrm{cm^{-3}}$ (Fig.~\ref{fig:sil_abund_cond}{\it  i}).
For $n_{\mathrm{H}} \gtrsim 10^{8} \ \mathrm{cm^{-3}}$, the contribution of dust to the Pedersen conductivity becomes small
as the small dust particles are depleted by the coagulation growth of dust particles (Fig.~\ref{fig:sil_size}{\it a}),
and the contribution from ions is the largest among the charged particles (Fig.~\ref{fig:sil_abund_cond}{\it i}).

When collisional fragmentation is included (model sil-frag), the contribution of electrons to Ohmic conductivity 
is smaller than that in model sil-coag (Fig.~\ref{fig:sil_abund_cond}{\it e} and {\it f}).  
This is because the abundance of electrons in model sil-frag is smaller than in model sil-coag (Fig.~\ref{fig:sil_abund_cond}{\it a} and {\it b}).
Even for model sil-frag, the total Ohmic conductivity is still dominated by electrons.
The Hall conductivity is larger in model sil-frag than in model sil-coag for $10^{6} < n_{\mathrm{H}} < 10^{11} \ \mathrm{cm^{-3}}$ (Fig.~\ref{fig:sil_abund_cond}{\it c} and {\it d}).
For model sil-frag, the contribution to $\sigma_{H}$ by ions is larger than that by electrons for $n_{\mathrm{H}} > 10^{6} \ \mathrm{cm^{-3}}$ (Fig.~\ref{fig:sil_abund_cond}{\it h}) 
because ions are more abundant than electrons (Fig.\ref{fig:sil_abund_cond}{\it b}).
For $n_{\mathrm{H}} > 10^{12} \ \mathrm{cm^{-3}}$, the contribution of ions to the Hall conductivity is sharply reduced, 
and the electron contribution is also small. 
As a result, the total Hall conductivity for model sil-frag is smaller than that for model sil-coag for this density range (Fig.~\ref{fig:sil_abund_cond}{\it c} and {\it d}).
For $10^{6} < n_{\mathrm{H}} < 10^{7} \ \mathrm{cm^{-3}}$, the contribution of dust to Hall conductivity is non-negligible relative 
to the total Hall conductivity (Fig.~\ref{fig:sil_abund_cond}{\it h}).
This trend is largely caused by the production of dust smaller than $0.1 \ \mathrm{\mu m}$ due to collisional fragmentation in the relatively low-density region (Fig.~\ref{fig:sil_size}{\it b}).
It is also related to the relatively large contribution of dust to the Pedersen conductivity compared to model sil-coag at relatively low densities (Fig.~\ref{fig:sil_abund_cond}{\it i} and {\it j}).
For $n_{\mathrm{H}} > 10^{12} \ \mathrm{cm^{-3}}$, $\sigma_{P}$ is smaller in model sil-frag than in model sil-coag (Fig.~\ref{fig:sil_abund_cond}{\it c} and {\it d})
because the abundance of ions is smaller in model sil-frag than in model sil-coag (Fig.~\ref{fig:sil_abund_cond}{\it a} and {\it b}).

The first row of Figure~\ref{fig:sil_res} shows the magnetic diffusion coefficients, $\eta_{O}, \eta_{H}$, and $\eta_{A}$ 
against the number density for models sil-coag (Fig.~\ref{fig:sil_res}{\it a}) and sil-frag (Fig.~\ref{fig:sil_res}{\it b}).
For model sil-coag (Fig.~\ref{fig:sil_res}{\it a}), the Ohmic diffusion coefficient $\eta_{O}$ is inversely proportional to the Ohmic conductivity $\sigma_{O}$ (eq.~\ref{eq:eta_ohm}) 
and monotonically increases as the density increases.
However, $\eta_{O}$ is smaller than $\eta_{H}$ and $\eta_{A}$ for the whole density range shown in Figure~\ref{fig:sil_res}{\it a}. 
In the low-density region of $ n_{\mathrm{H}} \lesssim 10^{6} \ \mathrm{cm^{-3}}$,  the ambipolar diffusion coefficient $\eta_{A}$ is larger than the other two coefficients.
$\eta_{A}$ gradually decreases as the density increases.
$\eta_{H}$ decreases slowly for $n_{\mathrm{H}} \lesssim 10^{6} \ \mathrm{cm}^{-3}$
and is almost constant in the range $10^{6} < n_{\mathrm{H}} < 10^{10} \ \mathrm{cm^{-3}}$.
The sign of $\eta_{H}$ changes from negative to positive at $n_{\mathrm{H}} \simeq 10^{10} \ \mathrm{cm^{-3}}$, corresponding to the sign change of $\sigma_{H}$ (Fig.~\ref{fig:sil_abund_cond}{\it c}).
In the region $n_{\mathrm{H}} > 10^{10} \ \mathrm{cm^{-3}}$, $\eta_{H}$ shows a slightly increasing trend and becomes the largest among the three diffusion coefficients, as shown  in Figure~\ref{fig:sil_res}{\it a}. 

Figure~\ref{fig:sil_res}{\it b} shows the case including collisional fragmentation (model sil-frag).
As in the case of model sil-coag, $\eta_{O}$ monotonically increases, and it is larger than that for model sil-coag
due to the decrease in $\sigma_{O}$ caused by the decrease in abundance of electrons (Fig.~\ref{fig:sil_abund_cond}{\it b}). 
$\eta_{A}$ monotonically increases for $n_{\mathrm{H}} > 10^{9} \ \mathrm{cm^{-3}}$ and becomes the largest for $n_{\mathrm{H}} > 10^{12} \ \mathrm{cm^{-3}}$ 
among the three coefficients. 
$\eta_{H}$ is almost constant in the range $10^{6} \mathrm{cm^{-3}} \lesssim n_{\mathrm{H}} \lesssim 10^{12} \ \mathrm{cm^{-3}}$.

To evaluate whether non-ideal MHD effects affect the dynamics, the magnetic Reynolds number $R_{\mathrm{m}} = V L / \eta$ 
is calculated using the magnetic diffusion coefficients.
The Jeans length $L_{\rm J}$ and free fall velocity $v_{\rm ff} = \sqrt{4\pi G L_{J}^{2} \rho_{g} /3}$ are used  as the typical length scale $L$
and typical velocity $V$, respectively \citep{2007ApJ...670.1198M}. 
It is considered that non-ideal MHD effects apply and 
the magnetic field is effectively decoupled from the gas when $R_{\mathrm{m}} \lesssim 1$.
Note, however, that the Hall effect is not a dissipative effect, and thus an evaluation in terms of the magnetic Reynolds number is not appropriate.
The second row of Figure~\ref{fig:sil_res} shows the magnetic Reynolds number.

For model sil-coag (Fig.~\ref{fig:sil_res}{\it c}), the magnetic Reynolds number never reaches less than unity even in the high-density region and is as large as $R_{\mathrm{m}}\gtrsim 10^{2}$.
In other words, non-ideal MHD effects,  especially of Ohmic dissipation and ambipolar diffusion, do not effectively apply for magnetic field diffusion.
On the other hand, for model sil-frag (Fig.~\ref{fig:sil_res}{\it d}), $R_{\mathrm{m}} \lesssim 1$ is realized for both ambipolar diffusion (in the range of  $n_{\mathrm{H}} \gtrsim 10^{12} \ \mathrm{cm^{-3}}$)  and Ohmic dissipation (in the range of  $n_{\mathrm{H}} > 10^{13} \ \mathrm{cm^{-3}}$). 
Therefore, when dust fragmentation is included, non-ideal MHD effects can contribute to the removal of the magnetic field in the high-density region.

\subsection{$\mathrm{H_{2}O}$ ice dust case}
\begin{figure*}
  \includegraphics[width=\linewidth]{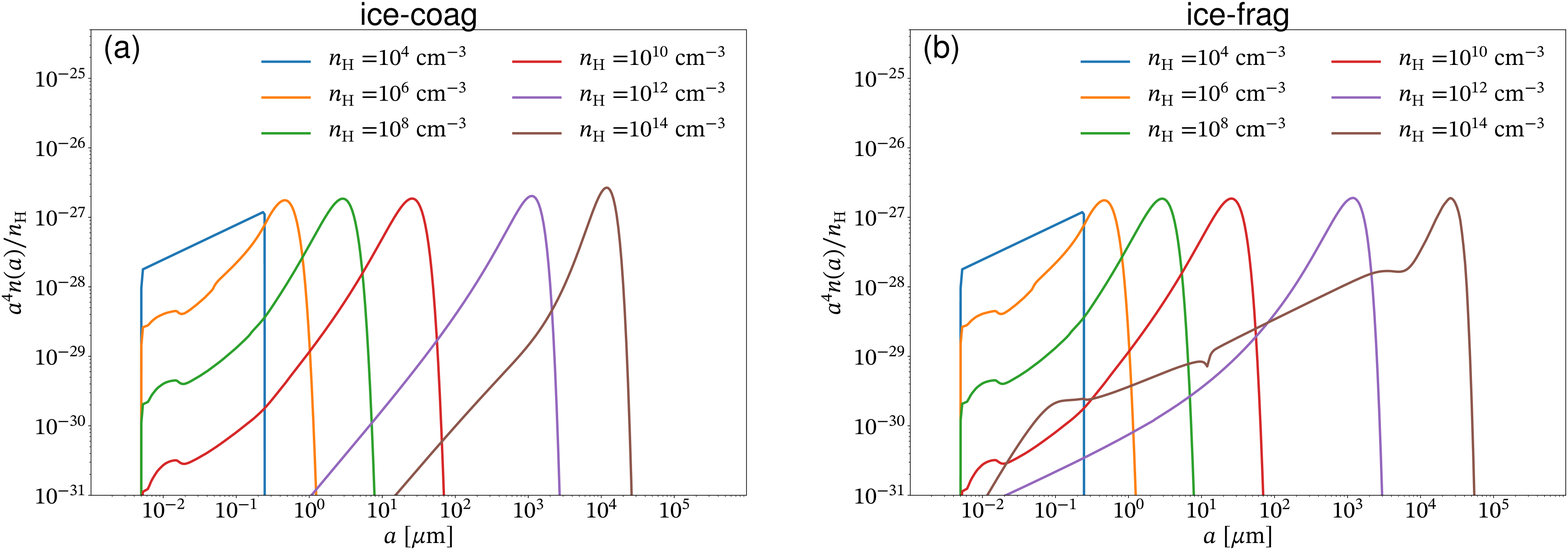}
  \caption{As Fig.~\ref{fig:sil_size} but for dust with a surface covered with $\mathrm{H_{2}O}$ ice (models ice-coag and ice-frag).}
  \label{fig:ice_size}
\end{figure*}

\begin{figure*}
  \includegraphics[width=140mm]{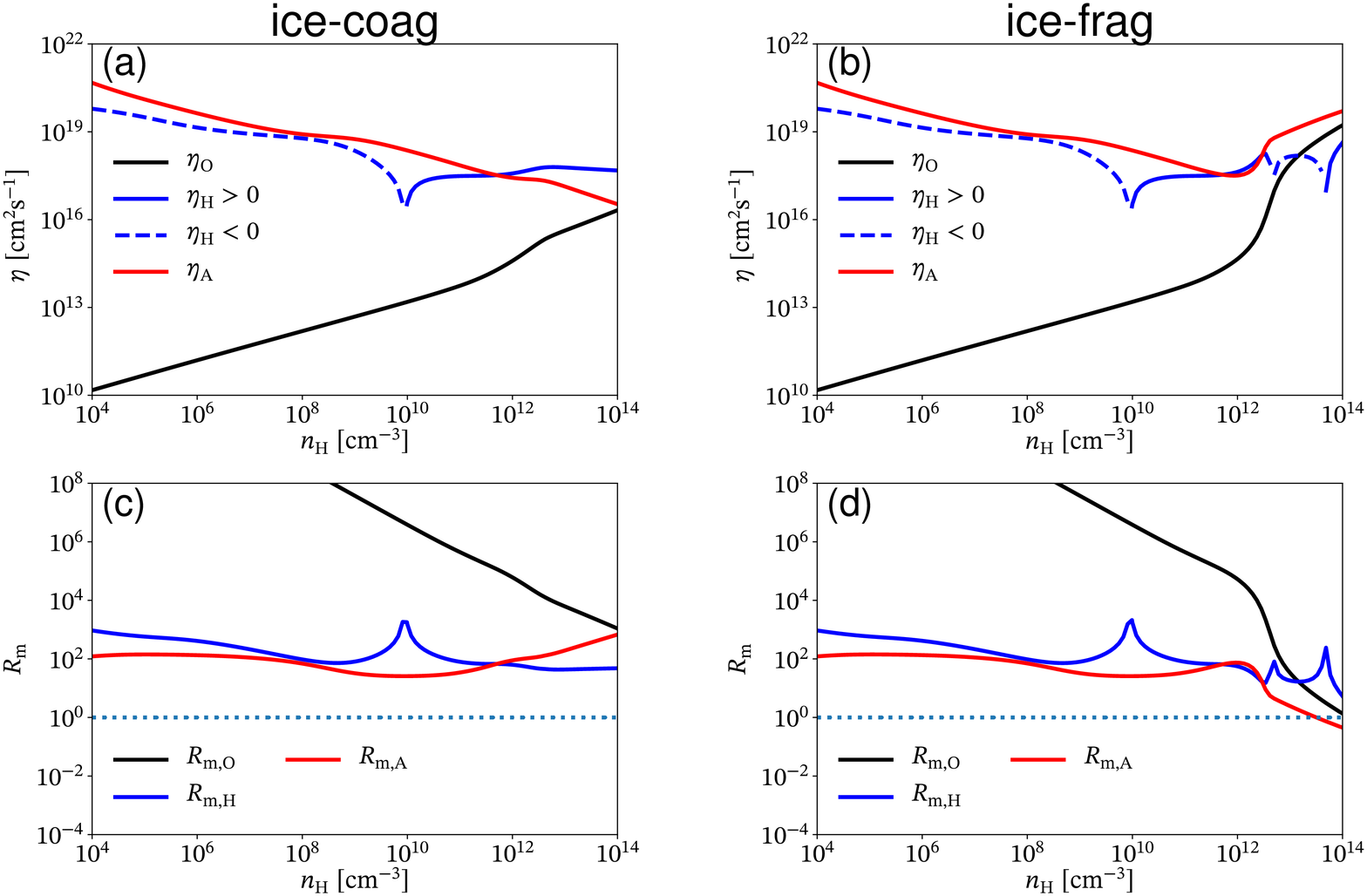}
  \caption{As Fig.~\ref{fig:sil_res} but for dust with a surface covered with $\mathrm{H_{2}O}$ ice (models ice-coag and ice-frag).}
  \label{fig:ice_res}
\end{figure*}

Figure~\ref{fig:ice_size} shows the evolution of the particle size distribution for dust with a surface covered with $\mathrm{H_{2}O}$ ice (models ice-coag and ice-frag).
In the case of $\mathrm{H_{2}O}$ ice, the threshold velocity required for collisional fragmentation $v_{f}$ is greater than that for silicate (Fig.~\ref{fig:rel_vel}).
Thus, even when collisional fragmentation is included, the effect of fragmentation
on the size distribution evolution is not significant for  $n_{\mathrm{H}} \lesssim 10^{12} \ \mathrm{cm^{-3}}$.
For $n_{\mathrm{H}} \gtrsim 10^{12} \ \mathrm{cm^{-3}}$, collisional fragmentation can produce small dust particles, 
and the particle size distribution evolution is different from the case with only coagulation growth (model ice-coag).
At all densities, the maximum dust particle size is larger in the $\mathrm{H_{2}O}$ case than in the silicate case. 
The dust grows to about $a \simeq 10^{4} \mathrm{\ \mu m}$ at $n_{\mathrm{H}}=10^{14} \ \mathrm{cm^{-3}}$ for the $\mathrm{H_{2}O}$ ice case for both ice-coag and ice-frag models.

As for models sil-coag and sil-frag, we calculated the abundance of charged particles and estimated the magnetic diffusion coefficients 
for models ice-coag and ice-frag.
For reference, the abundance of charged particles and electrical conductivities are presented in Appendix~\ref{appendix:H2O_ice_abund_cond}.
Figure~\ref{fig:ice_res} shows the magnetic diffusion coefficients (top panels) and the magnetic Reynolds number (bottom panels) for the $\mathrm{H_{2}O}$ ice case.
For $n_{\mathrm{H}} \lesssim 10^{12} \ \mathrm{cm^{-3}}$, the magnetic diffusion coefficients and magnetic Reynolds numbers have similar values 
for the models with and without collisional fragmentation.
For $n_{H} \gtrsim 10^{12} \ \mathrm{cm^{-3}}$, the magnetic diffusion coefficients are larger in model ice-frag than in model ice-coag.
In model ice-frag, the magnetic Reynolds number of ambipolar diffusion is below unity for $n_{\mathrm{H}}>10^{13} \ \mathrm{cm^{-3}}$, 
while for Ohmic dissipation it decreases below unity around $n_{\mathrm{H}}=10^{14} \ \mathrm{cm^{-3}}$.
Thus, for $\mathrm{H_{2}O}$ ice, magnetic field dissipation due to non-ideal MHD effects should occur only when collisional fragmentation is included, as for the silicate case.
However, the density range for which the condition $R_{\mathrm{m}}<1$ is realized is narrower  for model ice-frag than for model sil-frag. 

\section{discussion}
\label{sec:discussion}
\subsection{Magnetic braking catastrophe and magnetic flux problems}

\begin{figure}
  \includegraphics[width=\linewidth]{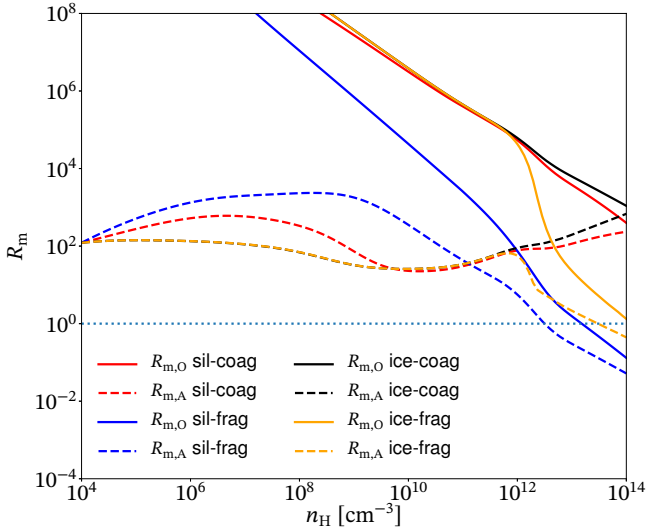}
  \caption{Magnetic Reynolds number for Ohmic dissipation and ambipolar diffusion against number density for all models.}
  \label{fig:magrey}
\end{figure}

There are two problems related to magnetic fields during star formation: the magnetic braking catastrophe and magnetic flux problems.
In the magnetic braking catastrophe, angular momentum is excessively transported due to very efficient magnetic braking during the star formation process, 
resulting in the failure of the formation of a circumstellar disk \citep{2003ApJ...599..363A,2008ApJ...681.1356M}.
Magnetic flux problems are cases where the magnitudes of the magnetic fluxes of the pre-contraction molecular cloud core (i.e. prestellar core) 
and the formed star differ over five orders of magnitude \citep{1953MNRAS.113..357B, 1984FCPh....9..139N,2002ApJ...573..199N,2020SSRv..216...43Z}.
These problems are considered to be solved by the dissipation of the magnetic field due to non-ideal MHD effects, 
in particular Ohmic dissipation and ambipolar diffusion.

Figure~\ref{fig:magrey} shows the magnetic Reynolds number for Ohmic dissipation and ambipolar diffusion estimated based on 
the dust particle size distribution evolution presented in this study for all models.
Non-ideal MHD effects (Ohmic dissipation and ambipolar diffusion) do not occur when only coagulation growth of dust is considered. 
Thus, if collisional fragmentation is not considered, the gas fluid approaches the ideal MHD regime and magnetic braking should be very effective, 
making disk formation more difficult. 
In addition, it is expected that the magnetic flux is not sufficiently removed from the center of the collapsing cloud core. 
These problems may be solved by considering both collisional fragmentation and  coagulation growth, 
because the removal of magnetic field can occur at high densities due to Ohmic dissipation and ambipolar diffusion.

Since this study is based on a one-zone model, three-dimensional non-ideal MHD simulations of two fluids composed of  dust and gas are necessary  to more precisely investigate the above two problems.
Although such simulations have been performed recently \citep{2019A&A...626A..96L,2020A&A...641A.112L,2021ApJ...913..148T,2021ApJ...920L..35T}, none have self-consistently  calculated dust growth and associated changes in the magnetic diffusion coefficients. 
Accurate estimates of the magnetic diffusion coefficients are needed to perform a simulation of the dust particle size distribution evolution.
Our results showed  that both dust coagulation and  fragmentation should be included for  an accurate estimate of the magnetic diffusion coefficients.
Three-dimensional simulations including both coagulation growth of dust and dust fragmentation are required to correctly understand the early star formation process, and these should be incorporated into future studies.

\subsection{Thermal evolution of star formation processes}
In this study, the density evolution of a collapsing cloud core is determined by a one-zone model, 
while the temperature is given by a barotropic equation of state, as described in \S\ref{sec:gascollapse}.

For a collapsing cloud core, \citet{2009MNRAS.399.1795H} solved the thermal evolution as well as the evolution of the dust particle size distribution due to coagulation using a one-zone model.
They concluded that the dust particle size distribution evolution has little effect on the thermal evolution of the collapsing cloud core.
However,  they only considered Brownian motion for the relative velocity of dust particles.
When the contribution of turbulence to the relative velocity is included, the evolution of the dust particle size distribution can be significantly changed compared to the calculation that only includes Brownian motion.
In addition, they  did not consider dust fragmentation.
As shown in this study, fragmentation also affects the dust particle size distribution evolution.
Thus, the thermal evolution may be  changed  when turbulence and dust fragmentation are considered.  
We will investigate the thermal evolution including the dust particle size distribution evolution presented in this study in future work.

\subsection{Turbulence model}
\label{sec:turbulence}
In this study, we used the turbulence model proposed by \citet{2007A&A...466..413O}, in which the fluctuating velocity of the largest turbulence eddy 
was assumed to be the sound speed that corresponds to a relatively strong turbulence.
As the relative velocity between dust particles due to turbulence increases,  the collision frequency also increases.  
The probability of collisional fragmentation then increases as the relative velocity increases.

The turbulence velocity in actual molecular cloud cores, which are in a quasi-static equilibrium state, have been found to be trans- or subsonic in observations 
\citep{2004A&A...416..191T,2007prpl.conf...63B}.
In contrast, in theoretical studies, 
it is possible to amplify the initially subsonic turbulence in a contracting core to be comparable to the speed of sound 
\citep{2021ApJ...915..107H,2021A&A...655A...3H}. 
When the magnitude of turbulence is reduced, the collision frequency  and probability of collisional fragmentation between dust particles decrease.
Therefore, it is considered that  the dust particles will coalesce and grow in size without showing frequent fragmentation  during the contraction phase of the cloud core. 
We will focus on the evolution of the dust particle size distribution for different turbulence strengths in future work.

\subsection{Fragmentation model}
In this study, the dust fragmentation model was obtained from the numerical calculations of dust collisions by \citet{2013A&A...559A..62W}.
\citet{2013A&A...559A..62W} showed that the ejected mass after a dust collision can be expressed by a simple relationship, as described in equation~(\ref{eq:eject_mass}).
\cite{2021ApJ...915...22H} also performed dust collision calculations with a very wide range of dust mass ratio 
and derived a more complex  relationship than equation (\ref{eq:eject_mass}).
However, these collision calculations were performed with dust consisting of a single-sized monomer.
Further sophisticated numerical collision calculations are needed because  the dust is considered to be composed of various sized monomers.

For the collisional fragmentation velocity, the velocity obtained from \citet{2013A&A...559A..62W} was also applied in this study.
However, the binding energy per unit mass was averaged over the size distribution to account for the effect of the initial dust particle size distribution (eq.~\ref{eq:average_dust_strength}).
There is a large uncertainty in the collisional fragmentation velocity, because some assumptions are required to derive this velocity, as described in Section~\ref{sec:fragmentation_model}.
The evolution of the dust particle size distribution and magnetic diffusion coefficients should differ, depending on the collisional fragmentation velocity.
When the collisional fragmentation velocity is large,  fragmentation is less likely to occur and non-ideal MHD effects are less likely to influence  the star and disk formation processes (Fig.~\ref{fig:magrey}).
On the other hand, the fact that disks around very young stars have been confirmed in observations \citep{Tobin2016,2017ApJ...834..178Y,2020ApJ...902..141S} suggests 
that the suppression of magnetic braking due to non-ideal MHD effects is significant. 
For star and disk formation processes, it may be possible to limit the collisional fragmentation velocity and the properties of the dust from observations.

\subsection{Accretion growth and evaporation}
Molecules such as H$_2$O and CO in the gas phase accrete on the dust surface, forming a mantle and increasing the dust particle size
\citep{2015ARA&A..53..541B}.
The evolution of the dust particle size distribution due to accretion is not accompanied by a change in the dust particle number density, but it increases the total surface area of the dust. 
\citet{2011MNRAS.416.1340H} showed that when the initial dust particle size obeys the MRN distribution, 
dust particles with a size of $\sim0.01 \ \mathrm{\mu m}$ increase in size due to accretion, affecting the minimum size of the distribution.
On the other hand, there is little effect of the distribution on the maximum size.
As described in Section~\ref{sec:fragmentation_model}, the binding energy between dust particles per unit mass increases as the size of dust particles decreases. 
Thus, dust growth due to accretion changes the minimum size of the dust particles, which may affect the collision velocity at which the particles begin to fragment.

In this study, we considered dust initially covered with $\mathrm{H_{2}O}$ ice instead of modeling  mantle formation.
When the temperature is as high as $T \gtrsim 150 \rm K$ in the high-density region, $\mathrm{H_{2}O}$ ice on the dust surface sublimates.
This breaks the bonds between monomers and may produce small dust particles.
In addition, the dust particle interior composed of silicate and other materials would be exposed as the $\mathrm{H}_{2}$O ice on the dust surface sublimates. 

Compared with $\mathrm{H_{2}O}$ ice, silicate is less likely to coagulate and grow. 
Thus, there can be a large number of small dust particles composed of silicate at high densities. 
Therefore, the non-ideal MHD effects may apply, as a result of the low ionization due to the efficient adsorption of electrons on the  surface of silicate dust.
Although mantle formation and sublimation were not considered in this study, they may be important for evaluating the magnetic diffusion coefficients.

\section{Summary}
\label{sec:summay}
In this study, we investigated the size distribution evolution of dust particles in a collapsing cloud core  to obtain the magnetic diffusion coefficients 
and evaluate the non-ideal MHD effects.
The density evolution of the collapsing cloud core is given by a one-zone model. 
We included not only collisional coagulation between dust particles but also collisional fragmentation, assuming that either the dust is composed of silicate or  the dust surface is covered with $\mathrm{H_{2}O}$ ice.

In the size distribution evolution of dust particles obtained from the models with only collisional coagulation, small dust particles are depleted by an increase in dust coagulation as the density increases.
Because of the reduced abundance of dust particles, the adsorption efficiency for charged particles (especially electrons) on the dust surface is lower.
Thus, electrons and ions are present to the same extent and the conductivity becomes large.
Therefore, the magnetic diffusion coefficients become small and the magnetic Reynolds number is always larger than unity even in the high-density region. 
As a result, magnetic field dissipation due to the non-ideal MHD effects should not be efficient.

When collisional fragmentation is included, silicate dust easily fragments even at low velocities, resulting in the generation of small dust particles. 
As the abundance of small dust particles increases, electrons are efficiently adsorbed on the dust surface and decrease in abundance, resulting in a decrease in conductivity and an increase in magnetic diffusion coefficients at high densities. 
In this case, the magnetic Reynolds number is below unity in the high-density region. 
Thus, the non-ideal MHD effects can work efficiently.
On the other hand, when the dust surface is covered with $\mathrm{H_{2}O}$ ice, the evolution of the dust particle size distribution is almost the same as when only collisional coagulation is considered 
up to $n_{\mathrm{H}}=10^{12} \ \mathrm{cm^{-3}}$ due to the fact that a high velocity is required for collisional fragmentation.
For $n_{\mathrm{H}}>10^{12} \ \mathrm{cm^{-3}}$, small dust particles are produced by collisional fragmentation and the dust abundance becomes high, 
resulting in a decrease in the abundance of charged particles and an increase in the magnetic diffusion coefficients. 
However, the magnetic Reynolds number falls below unity only in a narrow density range around $n_{\mathrm{H}} = 10^{14} \ \mathrm{cm^{-3}}$.

This study demonstrates that  it is necessary to include both collisional coagulation and collisional fragmentation when investigating the evolution of the dust particle size distribution and evaluating the  non-ideal MHD  effects in gravitationally collapsing cloud cores.

\section*{Acknowledgements}
The present study was supported by JSPS KAKENHI Grant (JP22J11129: YK, JP21H00046, JP21K03617: MNM).

\section*{Data Availability}
The data underlying this article are available on request.



\bibliographystyle{mnras}
\bibliography{example} 



\appendix
\section{Numerical implementation}
\label{appendix:numerical}

In this section, we describe  our numerical method for the dust coagulation-fragmentation equation (eq.[\ref{eq:coag-frag-equation}]).

To compute the evolution of the dust particle size distribution, the entire radius range (or mass range) is divided into $N_{\rm bin}$ bins.
The minimum and maximum sizes of dust in bin $i$ are represented by $a_{i-1/2}$ and $a_{i+1/2}$, and $a_{i-1/2} = a_{\mathrm{min}}\zeta^{i-1}$ 
and $a_{i+1/2} = a_{\mathrm{min}}\zeta^{i}$ are adopted, where $\zeta = \left(a_{\mathrm{max}}/a_{\mathrm{min}}\right)^{1/N_{\rm bin}}$.

The minimum mass in bin $i$ is $m_{i-1/2} = m_{\mathrm{min}}\eta^{i-1}$ 
and maximum mass is $m_{i+1/2} = m_{\mathrm{min}}\eta^{i}$, where $\eta = \left(m_{\mathrm{max}}/m_{\mathrm{min}}\right)^{1/N_{\rm bin}}$.
We then denote the dust radius and mass in each bin as $a_{i} = (a_{i-1/2}+a_{i+1/2})/2$ and $m_{i} = (4\pi/3)\rho_{s} a_{i}^{3}$.
In this study, the minimum radius is set to be $10^{-3} \ \mathrm{\mu m}$, 
the maximum radius is $10^{6} \ \mathrm{\mu m}$, and the number of divisions is set to $N_{\mathrm{bin}}=405$.

The size distribution in each bin is assumed to be constant except for the initial conditions, 
the mass density in bin $i$ is described by $\rho_{i}=\rho \left(m_{i}\right)\Delta m_{i}$, 
and the number density is $n_{i}=\rho_{i}/m_{i}$, where $\Delta m_{i} = m_{i+1/2}-m_{i-1/2}$.
The initial dust particle size follows the MRN distribution $\mathrm{d}n/\mathrm{d}a \propto a^{-q} \ \left( a_{\mathrm{min, ini}} < a < a_{\mathrm{max, ini}} \right)$.
Applying a power of $q = 3.5$, the minimum size is $a_{\mathrm{min,ini}}=5.0\times 10^{-3} \ \mathrm{\mu m}$, 
the maximum size is $a_{\mathrm{max, ini}}=2.5\times 10^{-1} \ \mathrm{\mu m}$, and
the initial density of bin $i$ is
\begin{equation}
  \rho_{i} = \rho_{g} f_{\mathrm{dg}} \frac{ a_{i+1/2}^{4-q} - a_{i-1/2}^{4-q} }{a_{\mathrm{max, ini}}^{4-q} - a_{\mathrm{min, ini}}^{4-q} }.
\end{equation}

Next we consider coagulation between two bins $(i, j, \ {\rm where} \  i \ge j)$. 
Given the existence of a mass range in each bin as described above, the dust mass created after coagulation is assumed to be distributed between the two bins $k$ and $k+1$. 
The bin $k$ is determined so that the condition $m_{k-1/2} < m_{i-1/2} + m_{j-1/2} < m_{k+1/2}$ is realized.
In this study, we estimate the fraction $f$ of the mass transferred to bin $k$ as follows,
\begin{equation}
  f = 
  \begin{cases}
    0 \ \ \ (m_{j} + m_{i-1/2} > m_{k+1/2}), \\
    1 \ \ \ (m_{j} + m_{i+1/2} < m_{k+1/2}), \\ 
    \frac{m_{k+1} - \left(m_{i} + m_{j}\right)}{m_{k+1} - m_{k}} \ \ \ (\mathrm{else}).
  \end{cases}
\end{equation}
The case $m_{j} + m_{i-1/2} > m_{k+1/2}$ corresponds to the situation in which  the mass produced by coagulation is too high for bin $k$.
On the other hand, the case $m_{j} + m_{i+1/2} < m_{k+1/2}$ is opposite to the case $m_{j} + m_{i-1/2} > m_{k+1/2}$.
In other cases, we adopt a fraction derived from mass conservation \citep{2008A&A...480..859B}, 
assuming that the mass $m_{i}+m_{j}$ is distributed to bin $k$ and bin $k+1$.

In the case of fragmentation, the masses of bins $i$ and $j$ are used to determine the mass distribution after fragmentation as described in Section~\ref{sec:fragmentation_model}.

The coagulation-fragmentation equation is calculated using an implicit scheme. 
The mass increase due to cloud collapse is solved by a simple Eulerian method, and the time step is taken to be 0.01 of the free-fall timescale.

The total mass of the dust must be conserved. 
However, if the difference between the minimum and maximum masses is very large and exceeds the accuracy of the numerical precision, 
it is difficult to maintain mass conservation \citep{2008A&A...480..859B}.
Therefore, in this study, when the relative error in mass exceeds a certain tolerance, the error is added back to the bins that existed initially
\citep{2013ApJ...764..146G}. 
This adjustment for the mass conservation has very little effect on the results.

\section{abundance and conductivities for H$_2$O ice}
\label{appendix:H2O_ice_abund_cond}

\begin{figure*}
  \includegraphics[width=140mm]{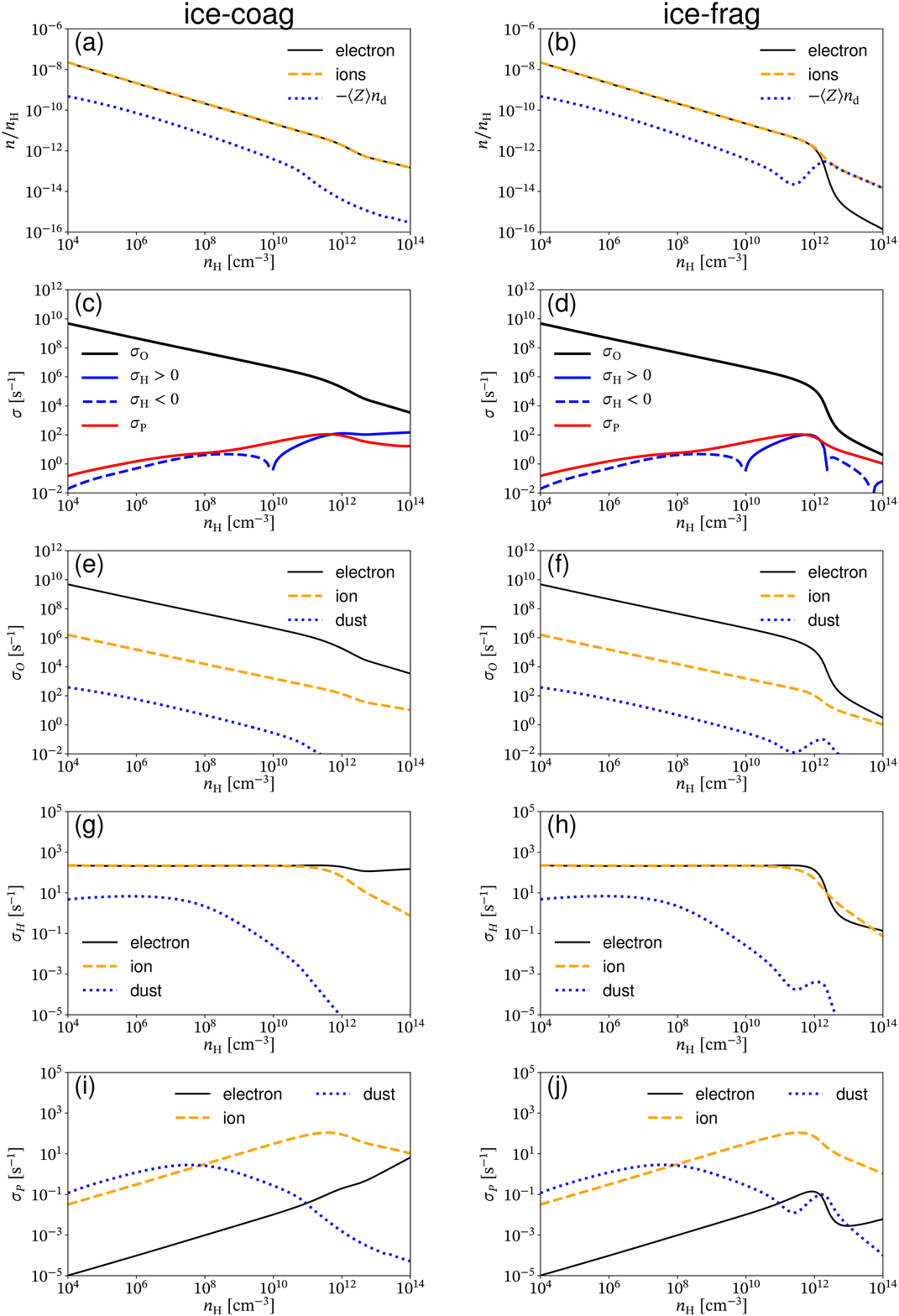}
  \caption{As Fig.~\ref{fig:sil_abund_cond} but for dust with a surface covered with $\mathrm{H_{2}O}$ ice (models ice-coag and ice-frag).}
  \label{fig:ice_abund_cond}
\end{figure*}

Figure~\ref{fig:ice_abund_cond} is the same as Figure~\ref{fig:sil_abund_cond} but for a dust surface covered with $\mathrm{H_{2}O}$ ice.
For model ice-coag, the electron and ion abundance are almost  the same (Fig.~\ref{fig:ice_abund_cond}{\it a})
due to the less efficient capture of electrons because of the reduced dust abundance compared to the silicate case.
The conductivity (Fig.~\ref{fig:ice_abund_cond}{\it c}) also shows the same behavior as in model sil-caog (Fig.~\ref{fig:sil_abund_cond}{\it c}).
However, the contribution of dust to the Hall and Pedersen conductivities in the range $10^{6} < n_{\mathrm{H}} < 10^{8} \ \mathrm{cm^{-3}}$ is 
smaller in model ice-coag than in model sil-coag  (Fig.~\ref{fig:ice_abund_cond}{\it g},{\it i} and Fig.~\ref{fig:sil_abund_cond}{\it g},{\it i})
because dust particles smaller than $0.1 \ \mathrm{\mu m}$ are depleted faster 
in the ice-coag case  than in the sil-coag case. 

When collisional fragmentation is included (model ice-frag), the dust particle size distribution in the density range $n_{\mathrm{H}} \lesssim 10^{12} \ \mathrm{cm^{-3}}$ 
is almost the same as that for model ice-coag (Fig.~\ref{fig:ice_size}).
Thus, the abundance of charged particles in this range is also  the same as that for model ice-coag (Fig.~\ref{fig:ice_abund_cond}{\it a} and {\it b}).
In the range $n_{\mathrm{H}} \gtrsim 10^{12} \ \mathrm{cm^{-3}}$,  collisional fragmentation produces small dust particles, increasing their abundance. 
As a result, the electron abundance is reduced due to more efficient electron adsorption on the dust surface and the net dust charge increases 
(Fig.~\ref{fig:ice_abund_cond}{\it b}).

The Ohmic conductivity for model ice-frag decreases for $n_{\mathrm{H}} > 10^{12} \ \mathrm{cm^{-3}}$ (Fig.~\ref{fig:ice_abund_cond}{\it d}), 
because the electron abundance shows a sharp decrease around  $n_{\mathrm{H}} \sim 10^{12} \ \mathrm{cm^{-3}}$ (Fig.~\ref{fig:ice_abund_cond}{\it b}).
For the Hall conductivity, in model ice-frag, the contribution of ions to $\sigma_{H}$ is larger than that of electrons 
around $n_{\mathrm{H}} \sim 10^{13} \ \mathrm{cm^{-3}}$ (Fig.~\ref{fig:ice_abund_cond}{\it h}), 
and the sign of $\sigma_{H}$ changes from positive to negative (Fig.~\ref{fig:ice_abund_cond}{\it d}).
For the Pedersen conductivity, the contributions from electrons and dust for model ice-frag differ significantly
from that for model ice-coag, especially in the range $n_{\mathrm{H}} > 10^{12} \ \mathrm{cm^{-3}}$ (Fig.~\ref{fig:ice_abund_cond}{\it i} and {\it j}). 
While the Pedersen conductivity $\sigma_{P}$ is dominated by the contribution from ions in both models (models ice-coag and ice-frag),
the value for model ice-frag is slightly smaller than that for model ice-coag (Fig.~\ref{fig:ice_abund_cond}{\it i} and {\it j}).
This is because the abundance of ions for model ice-frag is slightly smaller than that for model ice-coag (Fig.~\ref{fig:ice_abund_cond}{\it a} and {\it b}). 

\section{abundance and magnetic diffusion coefficients for a fixed dust size distribution (MRN size distribution)}
\label{appendix:MRN_no_size_evolution}

\begin{figure}
  \centering
  \includegraphics[width=70mm]{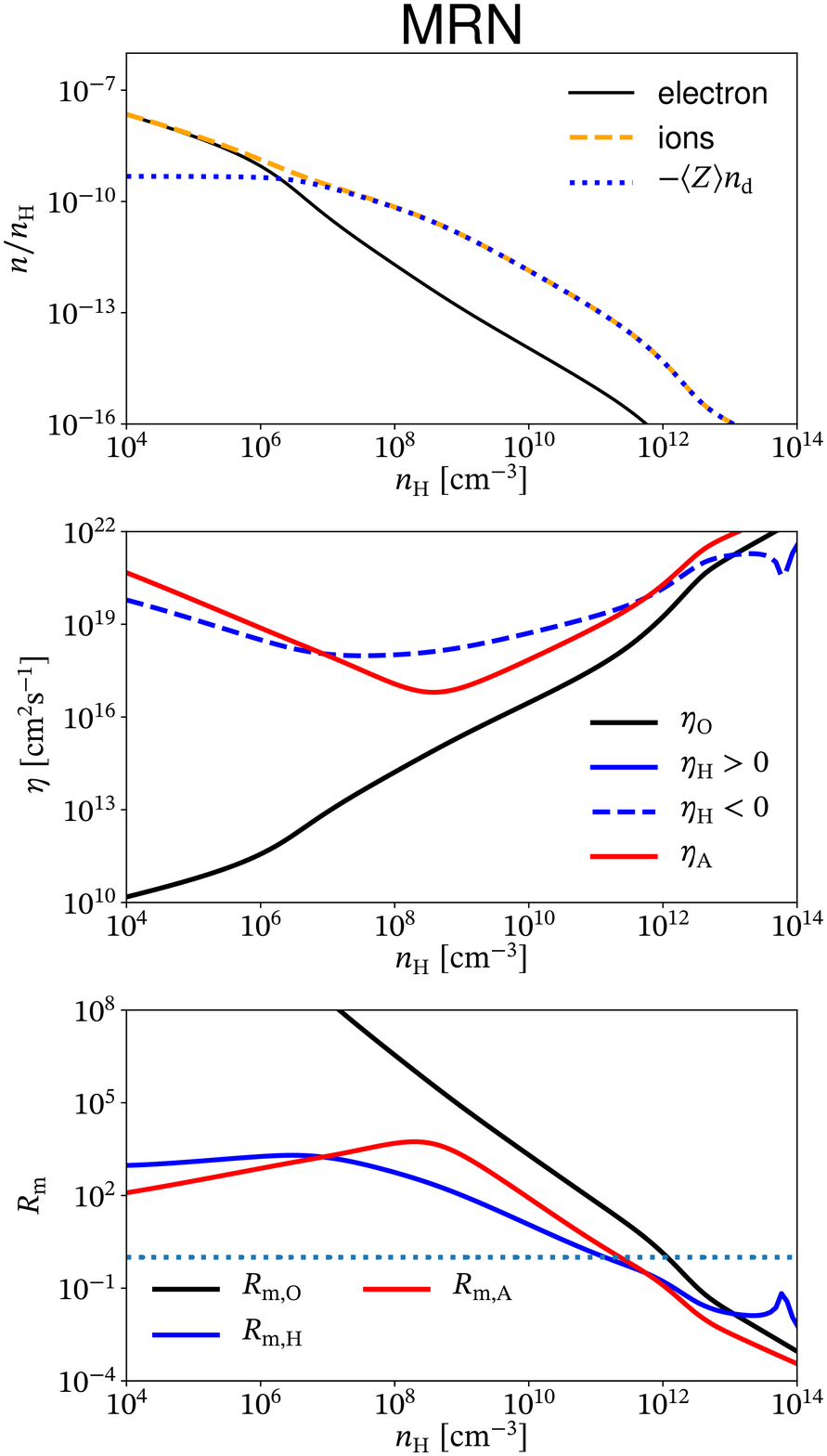}
  \caption{As Fig.~\ref{fig:sil_abund_cond} but for model MRN.}
  \label{fig:mrn_abund_eta}
\end{figure}

\begin{figure}
  \centering
  \includegraphics[width=70mm]{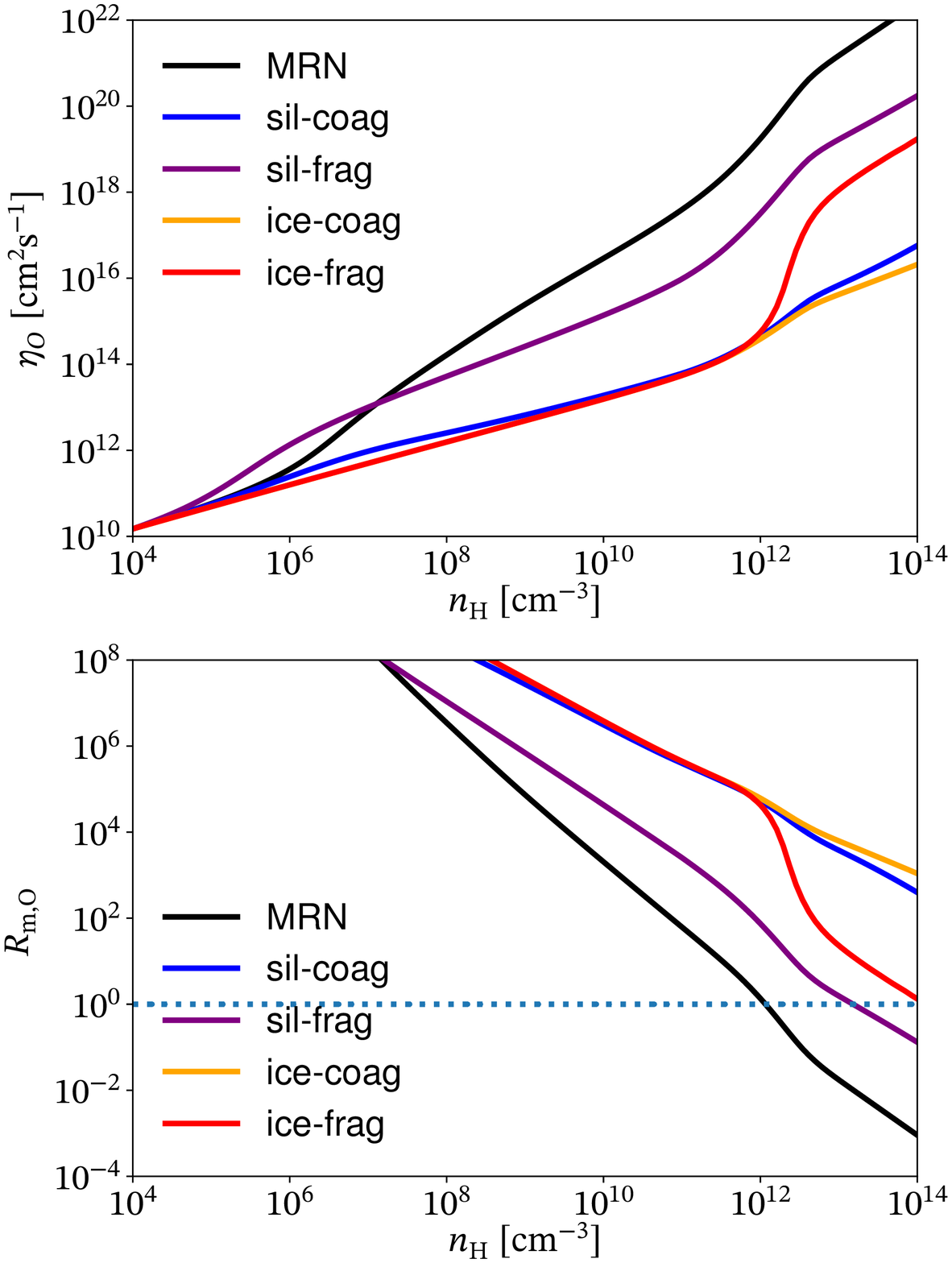}
  \caption{Comparison of the coefficient of Ohmic dissipation between the cases with (models sil-coag, sil-frag, ice-coag, and ice-frag) and without (model MRN) dust growth. 
}
  \label{fig:comp_res_ohm}
\end{figure}

\begin{figure}
  \centering
  \includegraphics[width=70mm]{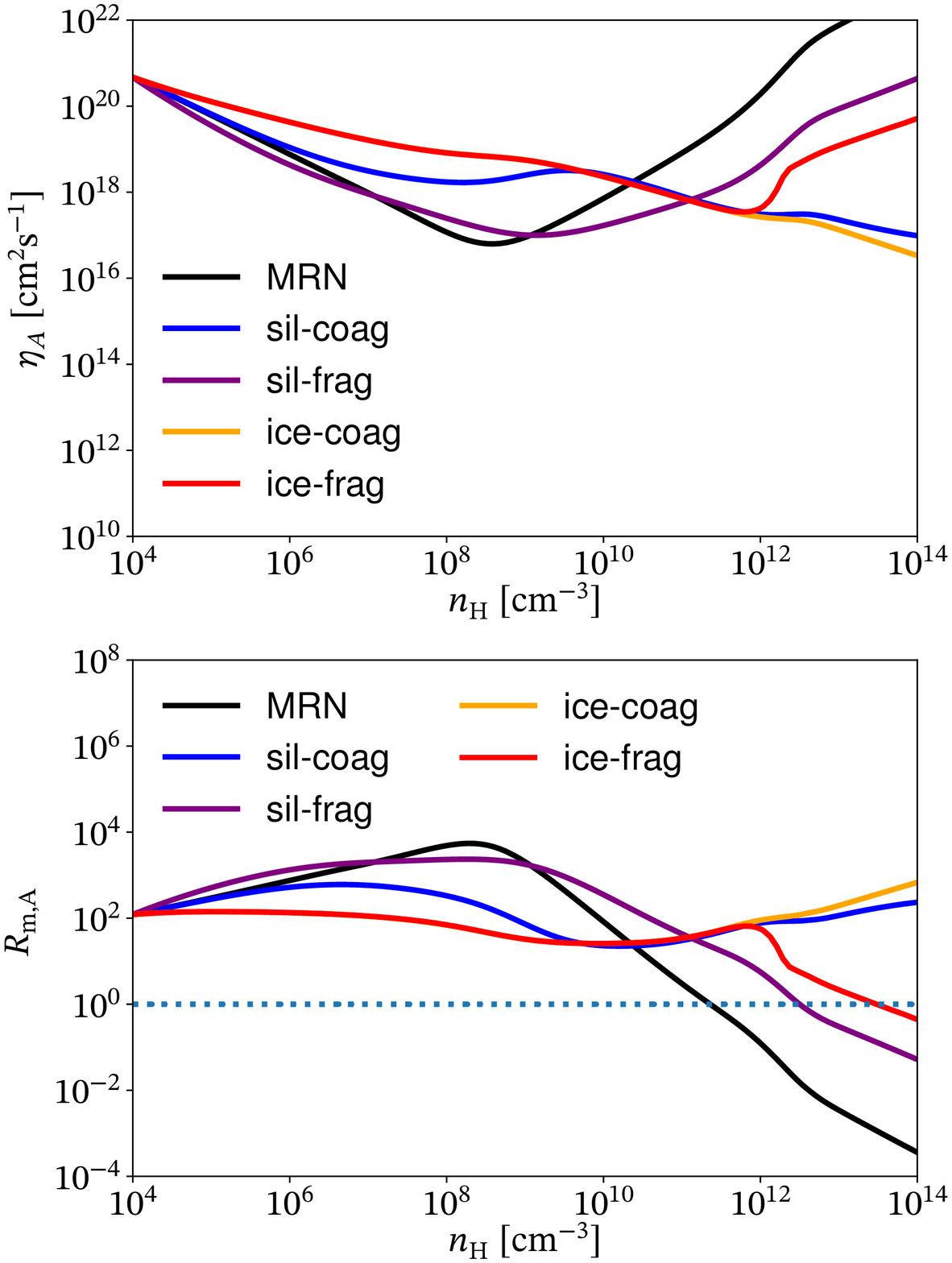}
  \caption{As Fig.~\ref{fig:comp_res_ohm} but for the coefficient of ambipolar diffusion.}
  \label{fig:comp_res_amb}
\end{figure}

We estimated the magnetic diffusion coefficients with a fixed dust size distribution  to compare the cases with and without dust growth. 
The procedure for the calculation is the same as that for the cases with dust growth described in \S\ref{sec:method}, but we adopted the MRN size distribution (see, \S \ref{sec:fragmentation_model}) without dust growth at all densities. 
We call the model without dust growth  `MRN' in this section. 

The calculation results for the MRN size distribution without dust growth (model MRN) are shown in Figure~\ref{fig:mrn_abund_eta}.
The top panel of Figure~\ref{fig:mrn_abund_eta} shows the abundance of charged particles. 
The electron and ion abundance are almost the same in the range  $n_{\mathrm{H}} \lesssim 10^{6} \ \mathrm{cm^{-3}}$, while 
efficient electron absorption on the dust surface reduces the abundance of electrons in the range $n_{\mathrm{H}} \gtrsim 10^{6} \ \mathrm{cm^{-3}}$.
The abundance of charged particles (electron, ion, charged dust particles)  for the case without dust growth (model MRN, top panel of Fig.~\ref{fig:mrn_abund_eta}) is much less than those for the cases with dust growth (models sil-coag, sil-frag, ice-coag, ice-frag,  top panels of Figs.~\ref{fig:sil_abund_cond} and \ref{fig:ice_abund_cond}).

The second panel of  Figure~\ref{fig:mrn_abund_eta} shows the magnetic diffusion coefficients against the number density for model MRN.
As the gas density increases, the Ohmic diffusion coefficient  $\eta_{O}$ monotonically increases. 
However,  $\eta_{O}$ is always smaller than  $\eta_{A}$.
The third panel of Figure~\ref{fig:mrn_abund_eta} shows that, for model MRN, the magnetic Reynolds number becomes less than unity $R_{\mathrm{m}} <1$  
for both ambipolar diffusion ($n_{\mathrm{H}} \gtrsim 10^{11} \ \mathrm{cm^{-3}}$) and Ohmic dissipation ($n_{\mathrm{H}} \gtrsim 10^{12} \ \mathrm{cm^{-3}}$).

Figure~\ref{fig:comp_res_ohm} shows a comparison of Ohmic dissipation between the case with (models sil-coag, sil-frag, ice-coag, ice-frag) and without (model MRN) dust growth, in which the coefficient $\eta_O$ (top) and magnetic Reynolds number $R_{\rm m,O}$ (bottom) of  Ohmic dissipation  are plotted.
The coefficient $\eta_{O}$ for model sil-frag is the largest among all models including model MRN in the range $n_{\mathrm{H}} \lesssim 10^{7} \ \mathrm{cm^{-3}}$.
On the other hand,  the coefficient $\eta_{O}$ for model MRN  is the largest in the range  $n_{\mathrm{H}} \gtrsim 10^{7} \ \mathrm{cm^{-3}}$.
Therefore, the  magnetic Reynolds number of Ohmic dissipation is below unity for model MRN  at a lower density than the other models.

Figure~\ref{fig:comp_res_amb} shows a comparison of ambipolar diffusion among all models.
In the range $n_{\mathrm{H}} \lesssim 10^{10} \ \mathrm{cm^{-3}}$, the coefficient $\eta_{A}$ for the models having less abundant small dust particles (models sil-coag, ice-coag) is larger than the other models  (models MRN, sil-frag, ice-frag).
The same trend can be seen in  \citet{2020A&A...643A..17G}. 
On the other hand, in the range  $n_{\mathrm{H}} \gtrsim 10^{10} \ \mathrm{cm^{-3}}$, the coefficient $\eta_{A}$ is larger for models having abundant small dust particles  (models MRN, sil-frag, ice-frag) than for models having less abundant small dust particles (models sil-coag,  ice-coag).
As well as Ohmic dissipation,  the magnetic Reynolds number of ambipolar diffusion becomes below unity for model MRN  at a lower density than the other models.

Finally, we comment on the method proposed by \citet{2021A&A...649A..50M}.
In this method,  the dust charge number is averaged for each dust particle size. 
The average charge number is used to determine the conductivities and the magnetic diffusion coefficients.
Although neutral dust particles are the most abundant in the range  $n_{\mathrm{H}} \gtrsim 10^{8} \ \mathrm{cm^{-3}}$, 
the abundance of charged dust particles is larger than that of electrons and ions \citep{2018MNRAS.478.2723Z,2019MNRAS.484.2119K}.
We confirmed that the contribution to the conductivities of the charged dust particles is somewhat underestimated (the magnetic diffusion coefficients are overestimated) with the averaged charge number method, especially when the average charge number of dust particles is small and nearly neutral. 

Furthermore, the mothod of \citet{2021A&A...649A..50M} does not take into account the charge transfer between dust particles. 
The charge transfer reduces the abundance of charged dust particles especially at high density of $n_{\mathrm{H}} \gtrsim 10^{12} \ \mathrm{cm^{-3}}$.
Thus, the inclusion of charge transfer should change the conductivities of dust particles in such a high-density region.
On the other hand, the contribution to the conductivities of dust particles is less significant compared with  that of electrons and ions when the abundance of small dust particles decreases due to coagulation. 
Thus, the effect of not considering the charge transfer is less  significant when dust growth is considered and the abundance of the small dust particles is relatively low. 
However, we would  update the method of \citet{2021A&A...649A..50M} to more accurately treat the dust charge number and the charge transfer.  


\bsp	
\label{lastpage}
\end{document}